\documentclass[12pt,preprint]{aastex}

\usepackage{lscape,longtable}
\shorttitle{Infrared study of \mbox{IRAS~10049-5657}}
\shortauthors{Vig et al.}
\begin{document}


\title{
Infrared study of the southern Galactic star forming regions associated with 
\mbox{IRAS~10049-5657} and \mbox{IRAS~10031-5632}
}

\author{S. Vig\thanks{Presently at INAF-Osservatorio Astrofisico di Arcetri, Largo E. Fermi 5, 50125 Firenze, Italy}, S. K. Ghosh, D. K. Ojha and R. P. Verma}
\affil{Tata Institute of Fundamental Research,
Mumbai (Bombay) 400 005, India}
\email{sarita@arcetri.astro.it}


\begin{abstract}
We investigate the physical conditions of the interstellar medium and 
stellar components in the regions of the southern Galactic star forming 
complexes associated with \mbox{IRAS~10049-5657} and \mbox{IRAS~10031-5632}. 
These regions  
have been mapped simultaneously in two far infrared bands 
($\lambda_{eff}$ $\sim$ 150 \& 210 $\mu$m), with $\sim 1'$ angular resolution 
using the TIFR 1-m balloon borne telescope. Spatial distribution of the 
temperature of cool dust and optical depth at 200 $\mu$m have been obtained 
taking advantage of the similar beams in the two bands. The HIRES processed 
IRAS maps at 12, 25, 60 and 100 $\mu$m have been used for comparison. 
Using the 2MASS near infrared sources, we find the stellar populations of the 
embedded young clusters. 
A rich cluster of OB stars is seen in the 
\mbox{IRAS~10049-5657} region.
The fits to the stellar density radial profile of the cluster 
associated with \mbox{IRAS~10049-5657} has been explored with inverse radius 
profile as well as the King's profile; the cluster radius is obtained to be 
$\sim2$ pc. The source in the cluster closest to the IRAS peak is 
IRA-7 which lies above the zero age main sequence curve of spectral type O5 
in the colour-magnitude diagram. 
Unlike \mbox{IRAS~10049-5657}, a small cluster comprising of a few 
deeply embedded sources is seen at the location of \mbox{IRAS~10031-5632}. Self 
consistent radiative transfer modelling aimed at extracting important physical 
and geometrical details of the two IRAS sources show that the best 
fit models are in good agreement with the observed spectral energy 
distributions. The geometric details of the associated cloud and optical depths
 ($\tau_{100}$) have been estimated. A uniform density distribution of dust 
and gas is implied for both the sources. In addition, the infrared 
ionic fine-structure line emission from gas has been modelled for both these 
regions and compared with data from IRAS-Low Resolution Spectrometer. For 
\mbox{IRAS~10049-5657}, the observed and modelled luminosities for most lines 
agree to within a factor of four while for \mbox{IRAS~10031-5632}, we find 
a discrepancy of a factor of 100 and it is likely that some 
basic assumptions of the model are not valid in this case.  
\end{abstract}



\section{Introduction}

HII regions are  astronomical sources that represent early stages of deeply
embedded high mass (O or early B) stars. Their study can provide vital
information about high mass star formation as well as their interaction with
the parent molecular cloud. Being deeply embedded in interstellar cloud 
including the dust component, almost all of their energy is absorbed and 
re-emitted in the infrared wavebands. \mbox{IRAS~10049-5657} (G282.0-1.2) and 
\mbox{IRAS~10031-5632} (G281.6-1.0) are Galactic star forming regions in the 
southern sky, which are generally less-studied. Radio 
measurements indicate that \mbox{IRAS~10049-5657} is an extended HII region 
(Hill, 1968; Manchester, 1969). 

A number of distance estimates to \mbox{IRAS~10049-5657} can be found in the
literature, ranging from 5.1 to 7.1 kpc. Here we use the distance of 6.3 kpc
estimated by Caswell \& Haynes (1987) based on radio recombination
line measurements (for R$_{\odot}$ = 8.5 kpc and Galactic rotation velocity =
220 km s$^{-1}$ at R$_{\odot}$). The distance to \mbox{IRAS~10031-5632} is 
estimated
to be 3.7 kpc (Caswell \& Haynes, 1987).

Recently, it has been concluded that \mbox{IRAS~10049-5657} complex harbours a 
very massive OB star cluster (Bik et al. 2005; Hanson et al., 2003). Bik et al. 
(2005) have carried out near-infrared K-band spectroscopy of few members 
of this cluster and find two very massive (O3-O4) stars here.
\mbox{IRAS~10049-5657} has been studied as a part of surveys for the 
search of emission lines including masers. Formaldehyde absorption has been 
detected towards this source at 4.8 GHz (Whiteoak and Gardner, 1974) and 
14.5 GHz (Gardner and Whiteoak, 1984). Whiteoak et al (1982) detected CO (1-0) 
line emission from this source using the 4-m radio telescope of CSIRO. Search 
for methanol transition (Peng and Whiteoak, 1992), methanol maser (Schutte et 
al, 1993) and OH maser (Cohen et al, 1995) close to this source have led to 
negative results. \mbox{IRAS~10049-5657} has been imaged by Puchalla et al. 
(2002) at 42 GHz using the Mobile Anisotropy Telescope on Cerro Toco (MAT/TOCO)
 and an integrated flux of $22.2\pm2.1$ Jy within a  
0.3$^{\circ}$ circular beam has been obtained by them. 
Both these regions, \mbox{IRAS~10049-5657} and \mbox{IRAS~10031-5632} have 
been studied as a part of the Parkes-MIT-NRAO
(PMN) survey (Kuchar and Clark, 1997). They find the peak radio flux density
of \mbox{IRAS~10049-5657} and \mbox{IRAS~10031-5632} at 5 GHz to be 25.7 
Jy/beam and 1.0 Jy/beam, respectively, where the beam is $\sim 4\arcmin.9$. 
CS~(2-1)
line emission has been observed close to both these regions (Bronfman et al., 
1996) using the SEST telescope. The IRAS-LRS (Low Resolution Spectrometer) 
spectrum of \mbox{IRAS~10031-5632} shows strong [Ne~II] emission line at 12.8 
$\mu$m (de Muizon et al., 1990; Simpson \& Rubin, 1990), a relatively weak 
[S~III] emission line at 18.7 $\mu$m as well as emission
 in the Unidentified Infrared Bands (UIBs) at 7.7, 8.6 and 11.3 $\micron$ 
(Zavagno et al., 1992; de Muizon et al., 1990). Water vapor maser (Braz et al. 
1989) as well as methanol maser (Schutte et al. 1993) have not been found close
 to \mbox{IRAS~10031-5632}.

The less known southern Galactic massive star forming regions are 
being studied under a long-term program which involves observing these sources 
in far infrared (Vig et al. 2007, Ojha et al. 2002, Karnik et al. 2001, 
Ghosh et al. 2000, Verma et al. 1994). 
In this paper, we present a systematic study of the star forming regions 
associated with \mbox{IRAS~10049-5657} and \mbox{IRAS~10031-5632}. The star 
forming region associated with \mbox{IRAS~10049-5657} is believed to harbour a 
cluster of very massive stars. \mbox{IRAS~10031-5632} is, on the other hand, a 
young 
star forming region with few members belonging to the cluster. We have carried 
out an infrared study of these southern Galactic HII regions in detail with 
the aim of understanding the energetics, 
physical sizes, the spatial distribution of interstellar dust and its 
temperature as well as the associated young clusters. Section 2 
describes observations and other data sets used. Section 3 describes the 
results and in Section 4, radiative transfer modelling of these sources is 
presented. A comprehensive discussion of these sources is carried out in 
Section 5 and a brief summary is presented in Section 6. 

\section{Observations and data reduction}

\subsection{Far infrared observations}
The Galactic star forming regions associated with \mbox{IRAS~10049-5657} and 
\mbox{IRAS~10031-5632} have been observed using the two-band far infrared (FIR) 
photometer system at the Cassegrain focus of 
the TIFR 100 cm (f/8) balloon borne telescope. The observations were carried 
out during the balloon flight from the TIFR Balloon Facility, Hyderabad in 
India (latitude $17^{\circ}.47$ north, longitude $78^{\circ}.57$ east) on Feb 
20, 1994. 
Details of the telescope and the observational procedure are given by Ghosh et
 al (1988). The two FIR bands use a pair of 2$\times$3 composite silicon 
bolometer arrays, cooled to 0.3K by liquid $^3$He which view 
identical parts of the sky simultaneously. The field of view of each detector 
is $1\arcmin.6$. The absolute positions were established from the detections of 
catalogued stars with an optical photometer located at the focal plane of the 
telescope, simultaneously during the FIR observations.
The planet Jupiter was observed at the beginning as well as at the
 end of the flight. 
The observations of Jupiter were used for the absolute flux 
calibration of the two band FIR photometer (12 channels) as well as for the 
determination of the Point Spread Function (PSF). 
The spectral response of each band of the FIR photometer was 
determined in the laboratory using a Michelson interferometer and a Golay cell 
as a comparison detector. The two FIR wavebands will be referred to 
as 150 and 210 $\micron$ bands corresponding to the $\lambda_{eff}$ for 30 K 
source with $\lambda^{-2}$ emissivity law. 

The region around \mbox{IRAS~10049-5657} ($\sim 32\arcmin \times 20\arcmin$) 
and \mbox{IRAS~10031-5632} ($\sim 25\arcmin \times 18\arcmin$) was mapped by raster 
scanning the region of the sky in cross-elevation with steps in elevation at 
the end of each scan. The FIR signals were gridded into a matrix with a pixel 
size of $0\arcmin.3 \times 0\arcmin.3$. The deconvolution of the observed 
chopped signal matrix was carried out using the Maximum Entropy Method similar 
to that of Gull \& Daniell (1978) (for details see Ghosh et al, 1988). 
An angular resolution of $\sim 1\arcmin$ has been achieved in the FIR maps 
using this method.  

\subsection{Other datasets used}
\subsubsection{IRAS}
The data from the Infrared Astronomical Satellite (IRAS) survey in the four 
bands (12, 25, 60 and 100 $\micron$)
for \mbox{IRAS~10049-5657} and \mbox{IRAS~10031-5632} were HIRES (High 
Resolution processing using Maximum Correlation Method; Aumann
et al., 1990) processed at the Infrared Processing and Analysis Center
(IPAC, Caltech) to obtain high angular resolution maps. 
The flux densities of the sources within a circular region of diameter 
$3'$ (centered on the peak) have been extracted from these images. 
\mbox{IRAS~10031-5632} appears in IRAS-LRS 
Catalog (IRAS Science Team, 1986) while the LRS spectrum of 
\mbox{IRAS~10049-5657} is 
presented by Volk \& Cohen (1989). 
These spectra, in the wavelength range 8-22 $\micron$, and the flux 
densities have been used for constructing the spectral energy distributions 
(SEDs).

\subsubsection{MSX}
The Midcourse Space Experiment\footnote{ This research made use of data
products from the Midcourse Space Experiment. Processing of the data was 
funded by the Ballistic
Missile Defense Organization with additional support from NASA Office of Space
Science. This research has also made use of the NASA/ IPAC Infrared Science
Archive, which is operated by the Jet Propulsion Laboratory, Caltech, under
contract with the NASA.} (MSX) surveyed the entire Galactic plane
within $|$b$|\le 5^{\circ}$ in four mid infrared wavebands: 8.3, 12.1, 14.7
and 21.3
$\micron$ at a spatial resolution of $\sim18\arcsec.3$ (Price et al., 2001).
The panoramic images of the Galactic plane
survey of MSX were taken from IPAC. 
The images of \mbox{IRAS~10049-5657} and \mbox{IRAS~10031-5632} were used to 
extract the sources and obtain the flux densities within a circular region of 
diameter $3'$ in order to construct the SEDs. Point sources close to 
these star forming regions have been selected from MSX Point Source Catalog 
Version 2.3 (Egan et al., 2003) and cross-correlated with 2MASS sources (see 
Sect. 2.2.3).

\subsubsection{2MASS}
The point sources around the regions \mbox{IRAS~10049-5657} and 
\mbox{IRAS~10031-5632} were
selected from the Two Micron All Sky Survey\footnote{This publication makes use of data products from the Two Micron All Sky
Survey, which is a joint project of the University of Massachusetts and the
Infrared Processing and Analysis Center/California Institute of Technology,
funded by the NASA and the NSF.} (2MASS) Point Source Catalog
(PSC).
The 2MASS PSC is complete down to J $\le15.8$, H $\le15.1$ and K$_s\le14.3$
mag for S/N$>10$, in the absence of confusion. The 2MASS sources used in this 
study are those with good photometric quality (rdflg $= 1-3$). The J, H and
K$_s$ magnitudes of the selected sources have been used to construct 
colour-magnitude (CM) and colour-colour (CC) diagrams which have 
been used to study the 
embedded clusters in these regions. The
JHK$_s$ magnitudes and images were taken from IPAC.

\subsubsection{SUMSS}
The Sydney University Molonglo Sky Survey (SUMSS) is a radio imaging survey of 
the sky south of declination $-30^{\circ}$ (Bock et al., 1999). This survey 
uses the Molonglo Observatory Synthesis Telescope\footnote{The MOST is 
operated by the University of Sydney and supported in parts by grants from the 
Australian Research Council.} (MOST), operating at 843 MHz with a 
bandwidth of 
3 MHz. The synthesised beam size is $\sim43''\times43''$cosec$|\delta|$.
The SUMSS radio images of \mbox{IRAS~10049-5657} and \mbox{IRAS~10031-5632} 
have been used to study the distribution of ionised gas around this region. 

\section{Results}

In this section, we present the results obtained from the observations as 
well as from the available data. As 
our study includes a wide wavelength range (near infrared to radio), we 
classify our results according to those for the dust and gas components 
(interstellar medium) versus the stellar component for both the regions, 
\mbox{IRAS~10049-5657} and \mbox{IRAS~10031-5632}.

\subsection{IRAS~10049-5657}

\subsubsection{Interstellar medium}
\begin {figure}
\epsscale{1}
\plottwo{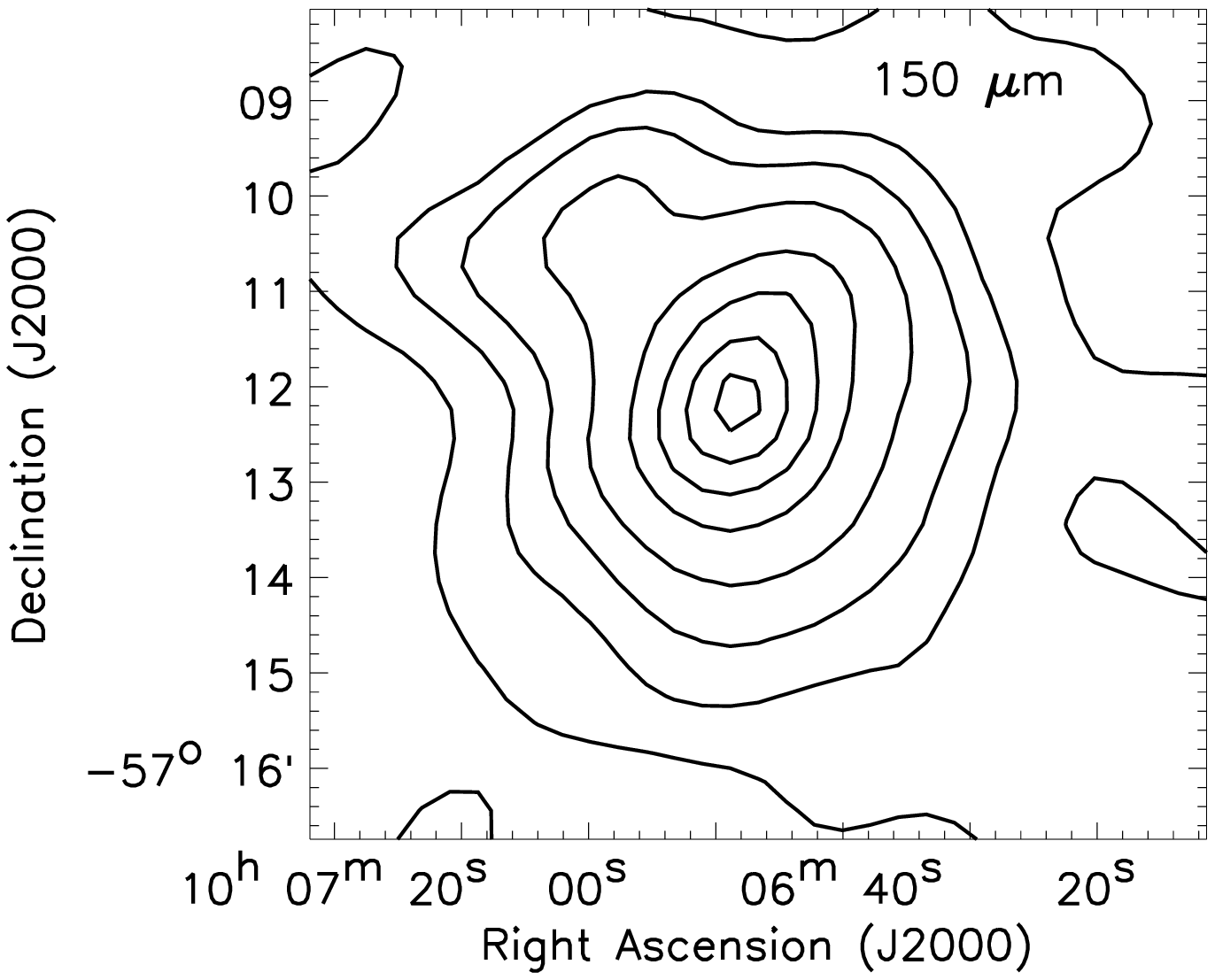}{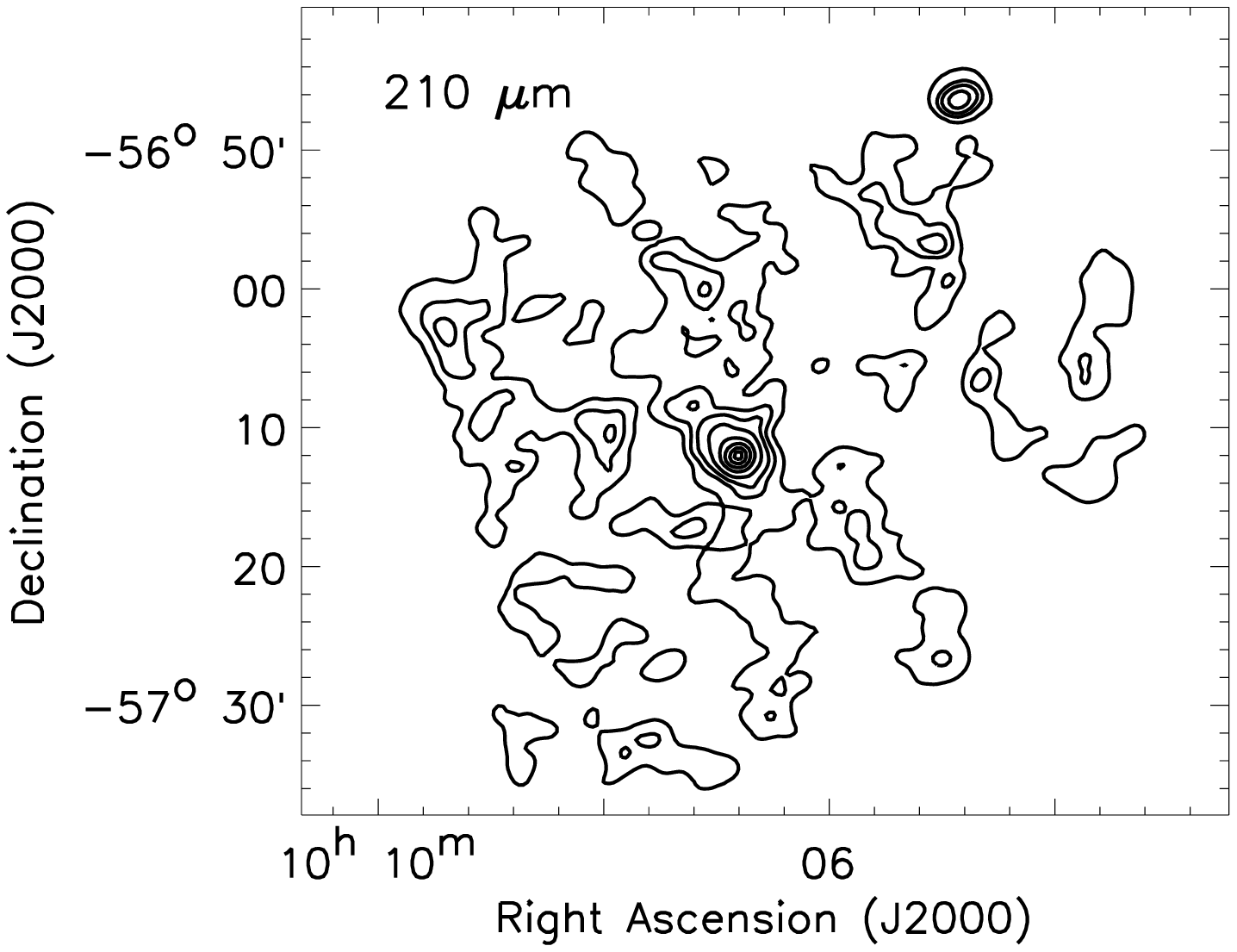}
\caption{The intensity map for the region around \mbox{IRAS~10049-5657} at 150 $\mu$m 
(left) and 210 $\mu$m (right). While the 150 $\mu$m map shows emission 
from \mbox{IRAS~10049-5657} region, the 210 $\mu$m map shows the complete region 
scanned by the far-infrared telescope. Contour levels are at 1, 5, 10, 20, 40,
 60, 80, 95\% of peak intensity of 624 Jy/sq arc min (left) and 292 Jy/sq arc 
min (right).
}
\label{FIRmap_a}
\end {figure}

\begin {figure}
\epsscale{1}
\plottwo{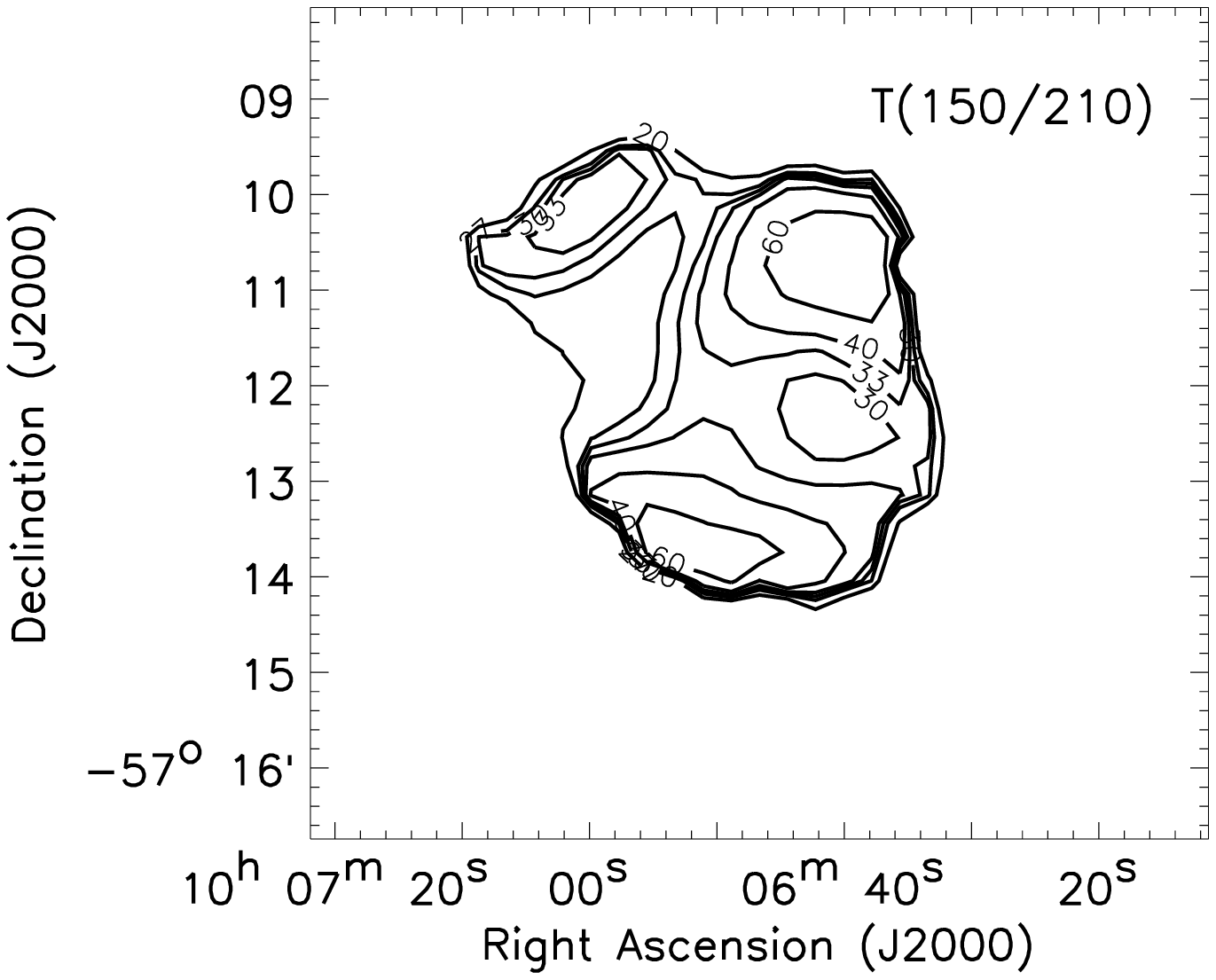}{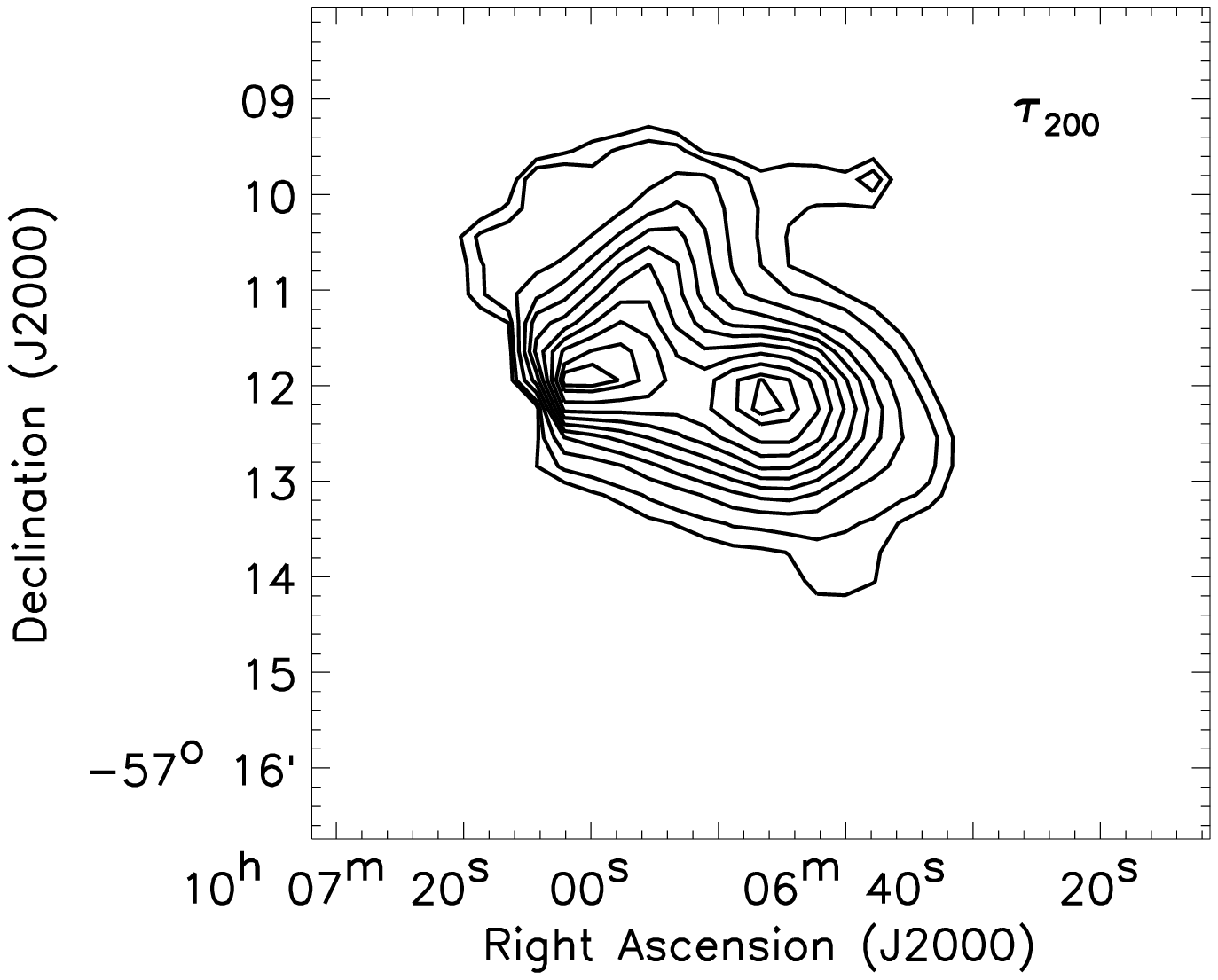}
\caption{The distribution of dust temperature T(150/210) (left), and
optical depth at 200 $\mu$m, $\tau_{200}$, (right) from the region around 
\mbox{IRAS~10049-5657} assuming a dust emissivity law of $\epsilon_{\lambda} \propto
\lambda^{-2}$. The isotherms of T(150/210) correspond to 20, 27, 30, 33, 
40 and 60 K. The
$\tau_{200}$ contours represent  5, 10,  20, 30, 40, 50, 60, 70, 80 and 90 \% of
 the peak value of $0.007$.
}
\label{Ttaubal_a}
\end {figure}

\begin {figure}
\plottwo{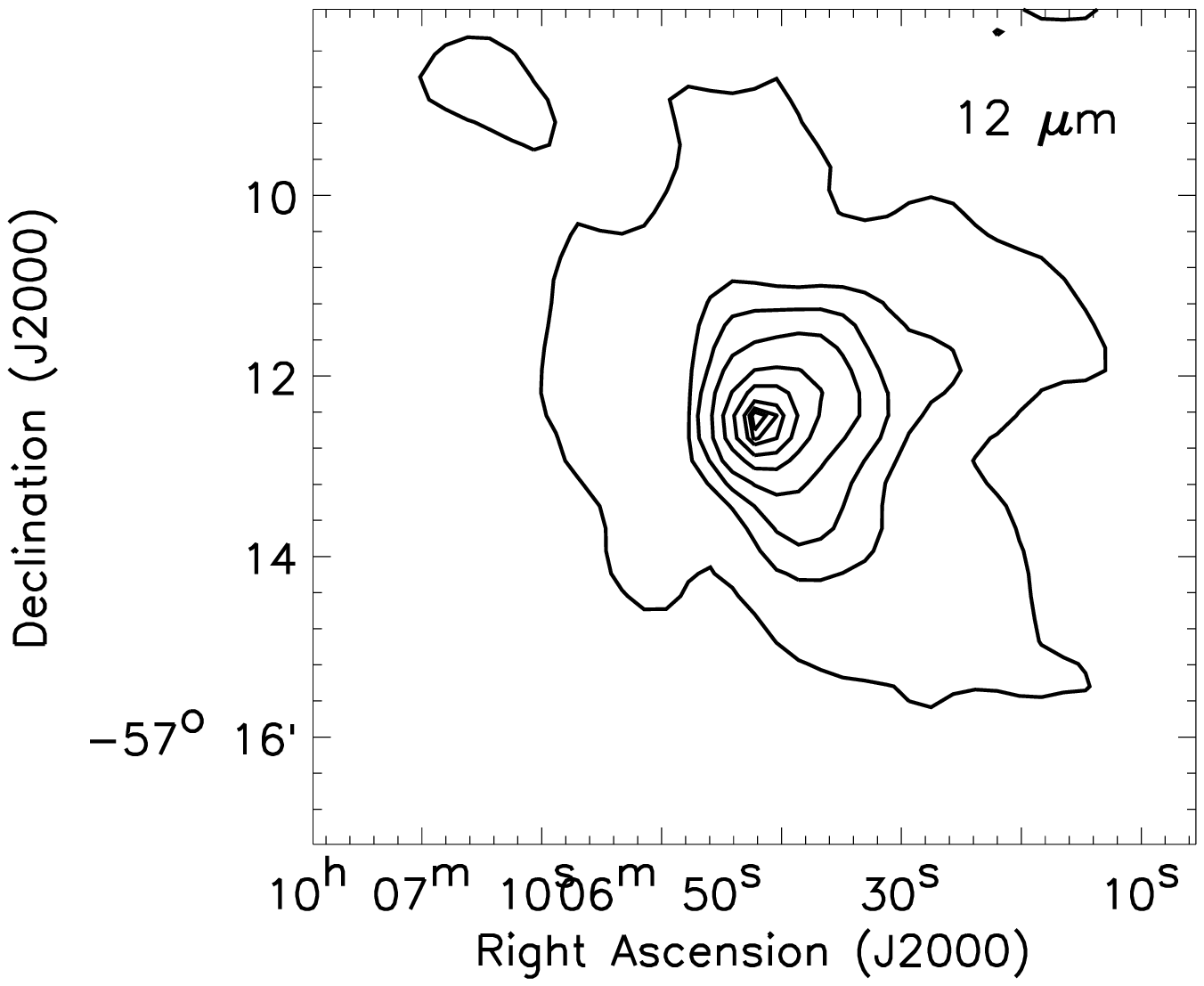}{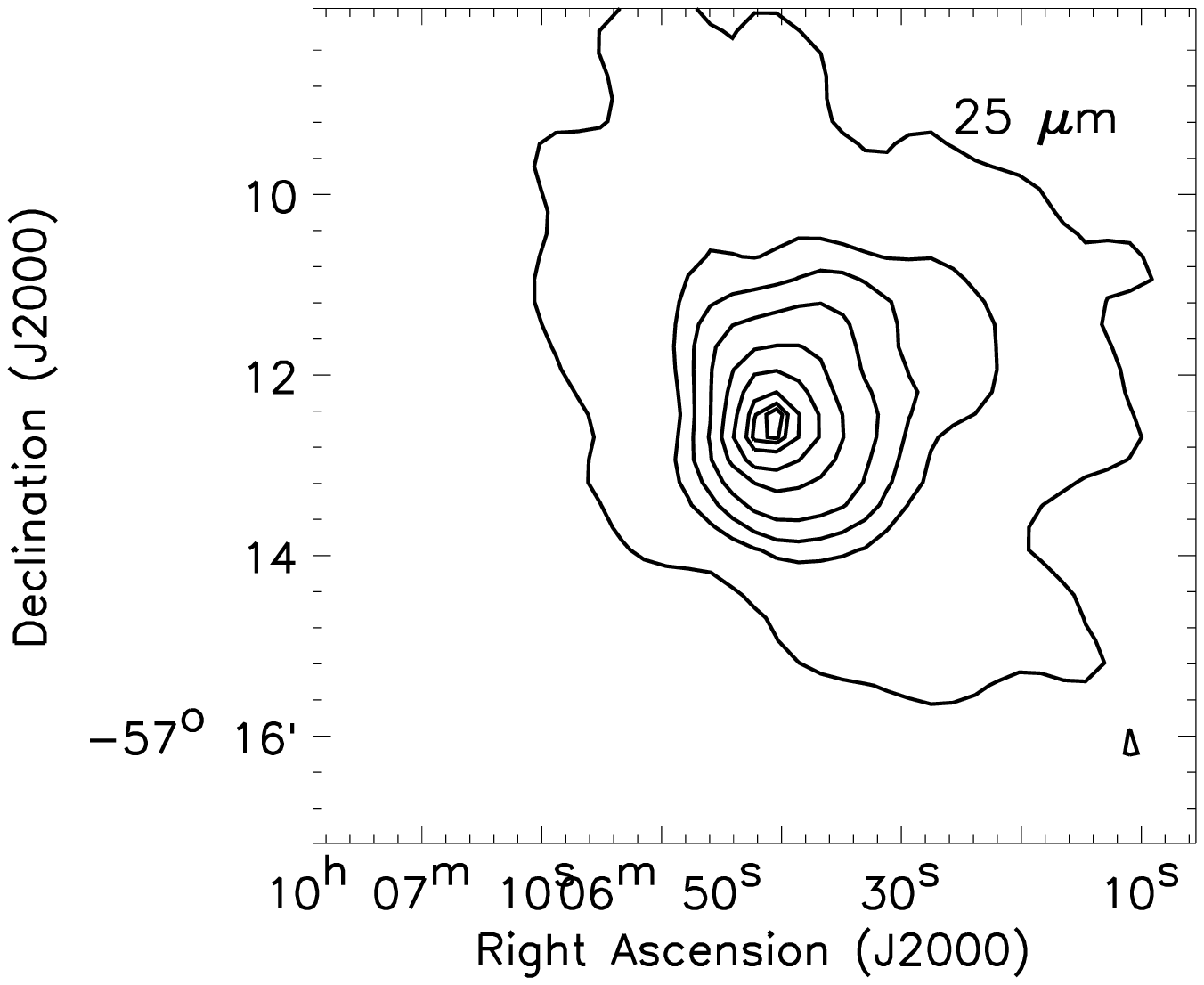}
\plottwo{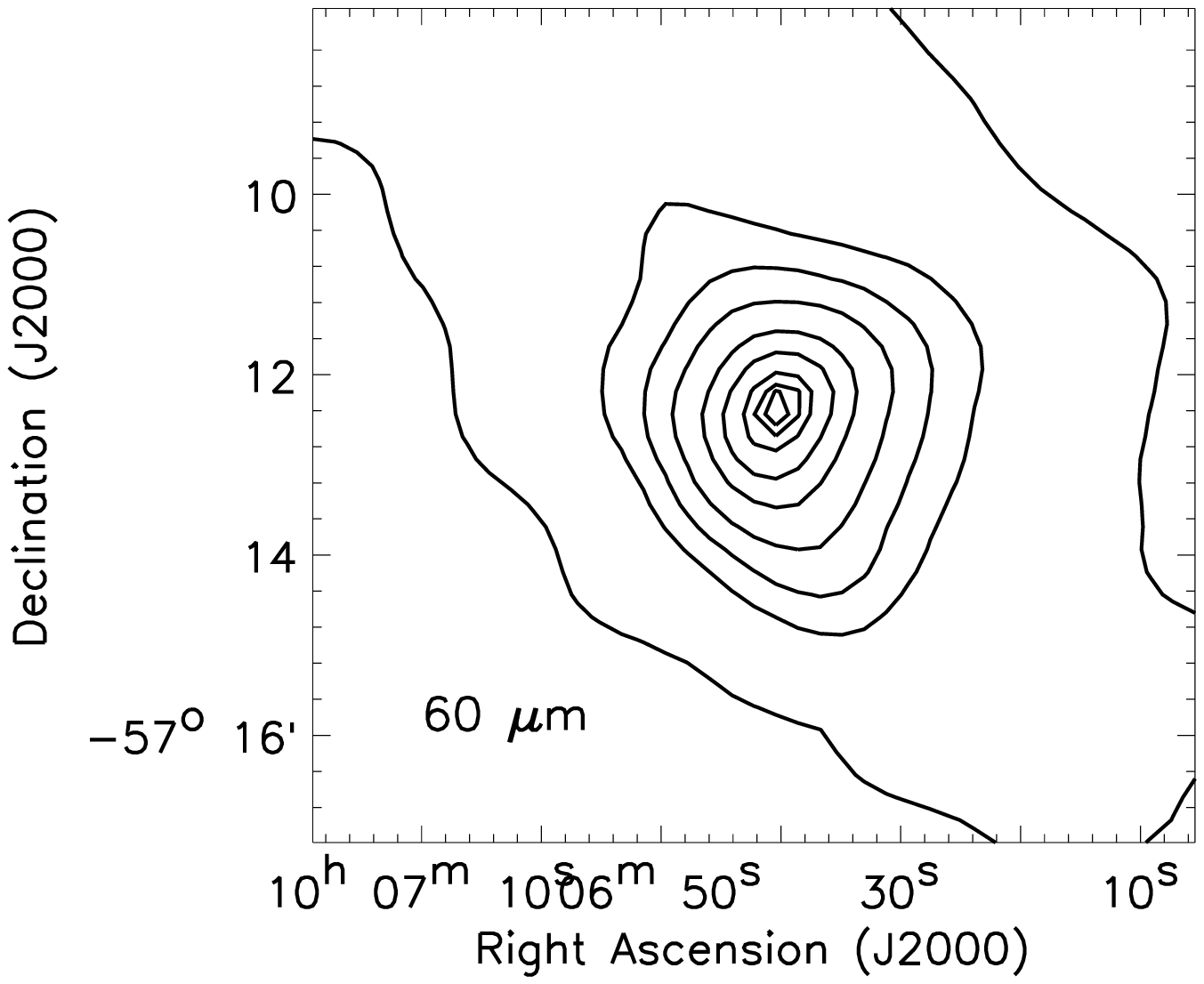}{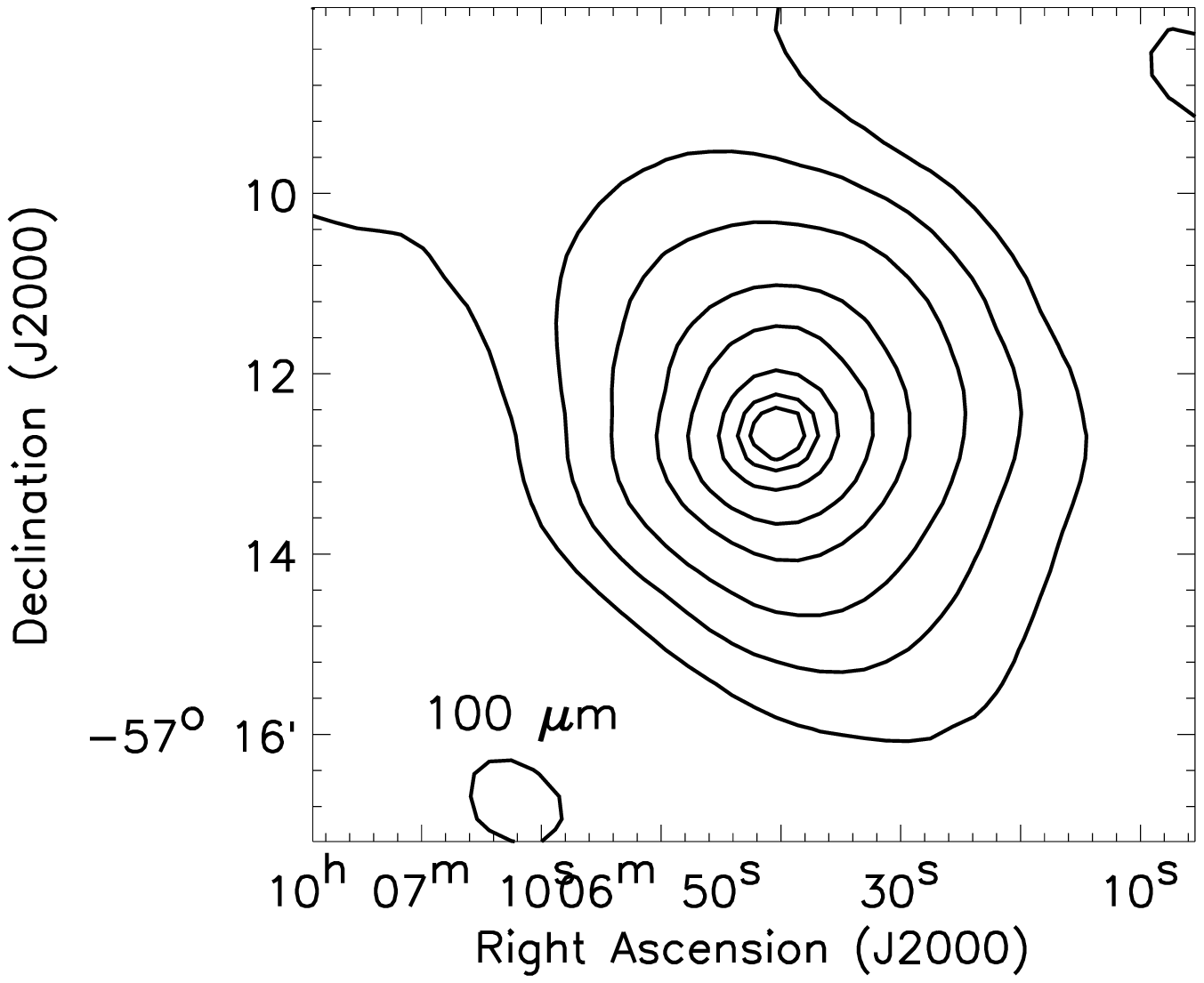}
\caption{The IRAS-HIRES intensity maps for the region covering 
\mbox{IRAS~10049-5657}
at 12 $\micron$ (top left), 25 $\micron$ (top right), 60 $\micron$ (bottom
left) and 100 $\micron$ (bottom right). The contours are at 1, 5, 10, 20, 40, 
60, 80, 90 and 95 \% of the peak value of 272 Jy arcmin$^{-2}$, 1053 Jy 
arcmin$^{-2}$, 3367 Jy arcmin$^{-2}$ and 1423 Jy
arcmin$^{-2}$ at 12, 25, 60 and 100 $\mu$m, respectively.
}
\label{HIRES_a}
\end {figure}

The deconvolved maps of \mbox{IRAS~10049-5657} at 150 and 210 $\mu$m are
presented in Fig.~\ref{FIRmap_a}. Figure~\ref{FIRmap_a} (left) shows  
emission from the \mbox{IRAS~10049-5657} region at 150 $\mu$m while 
Fig.~\ref{FIRmap_a} 
(right) shows the emission at 210 $\mu$m, from the complete region 
($32'\times20'$) mapped by the telescope. The angular size (50\% contour 
level) of \mbox{IRAS~10049-5657} at 150 and 210 $\mu$m is $\sim2'$. The FIR 
emission in these maps samples the cold dust around these regions. 
\mbox{IRAS~10049-5657} is well resolved in both the bands. The FIR emission in 
this region shows extension towards north-east in both the bands. This emission 
towards the north-east extends to a larger scale in the 210 $\mu$m map in 
Fig.~\ref{FIRmap_a} (right). Taking advantage of the simultaneous observations 
in the two bands, with almost identical field of view, we have generated maps 
of the dust temperature ($T_{\rm d}$) and optical depth at 200 $\mu$m 
($\tau _{200}$). The dust temperature T(150/210) and optical depth 
($\tau_{200}$) maps are shown in Fig.~\ref{Ttaubal_a}. For these maps, we 
have assumed dust emissivity of the form $\epsilon_{\lambda} \propto 
\lambda^{-2}$. The temperature distribution shows plateaus of maximum (60 K) 
towards north-west and south-east of the FIR peak. We have detected dust as 
cold as 20 K. The peak optical depth at 200 $\mu$m is determined to
be $0.007$ and is located at the position of the peak of FIR emission. 
Another peak is seen to the west of the maximum optical depth with an extension 
corresponding to north-east.
The IRAS-HIRES maps of the region around \mbox{IRAS~10049-5657} at 12, 25, 60 
and 100 
$\mu$m are shown in Fig.~\ref{HIRES_a}. The achieved angular resolutions are 
$0\arcmin.7\times0\arcmin.5$ at 12 and 25 \micron, 
$1\arcmin.4\times1\arcmin.1$ at 60 $\micron$, $2\arcmin.1\times1\arcmin.9$ at 
100 $\micron$. The flux densities of \mbox{IRAS~10049-5657} integrated over $3'$ 
diameter, centered on the peak, from the TIFR and IRAS-HIRES maps are 
listed in Table~\ref{fluxes}.  

\begin {figure*}
\hskip 3cm
\includegraphics[height=8.0cm]{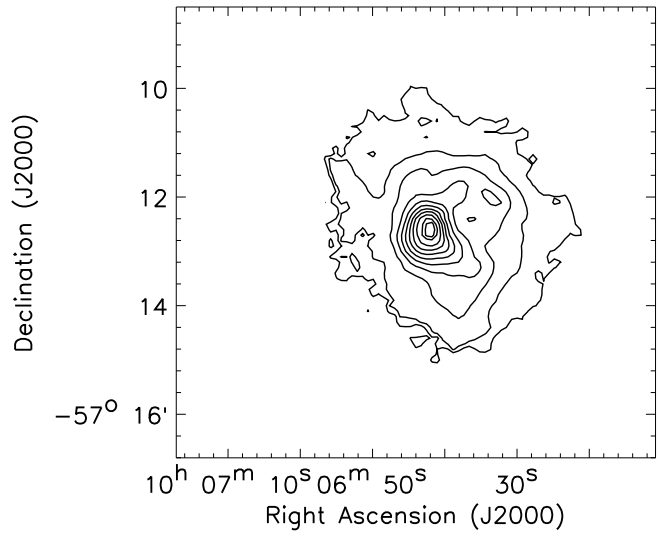}
\caption{The spatial distribution of total radiance in UIBs for the region 
around \mbox{IRAS~10049-5657} modelled using the MSX images. The
contour levels are at 5, 10, 20, 30, 40, 50, 60, 65, 70, 80, 90, 95\% of
peak value of $8.9\times10^{-5}$ W m$^{-2}$ Sr$^{-1}$.
}
\label{1004_pah}
\end {figure*}

We have modelled thermal continuum emission from interstellar dust alongwith
the emission in Unidentified Infrared Bands (UIBs), using mid-IR data from the 
MSX Galactic Plane Survey in the 8.3, 12.1, 14.7 and 21.3 $\micron$ bands. This 
has been carried out using the scheme developed by Ghosh \& Ojha (2002). In 
this scheme, the emission from  each ($6\arcsec \times6\arcsec$) pixel in the 
MSX images is 
modelled to be a combination of two components: (i) thermal continuum from 
the warm dust grains (gray body) and (ii) the emission from the UIB 
features falling within the MSX band. The scheme assumes that dust emissivity 
follows the power law of the form $\epsilon_{\lambda}\propto\lambda^{-1}$ and 
the total radiance due to UIBs in the 12 $\mu$m band is proportional to that 
in the 8 $\mu$m band. The spatial distribution of UIB emission 
predicted by this scheme is presented in 
Fig.~\ref{1004_pah}. The morphology of UIB emission shows extensions towards
the south-west near the peak as well as due north-west at the 
fainter levels. The peak strength of the modelled UIB emission is 
$8.9\times10^{-5}$ W m$^{-2}$ Sr$^{-1}$ and is close ($\sim16''$) to the IRAS 
position. Table~\ref{fluxes} also lists the flux densities of 
\mbox{IRAS~10049-5657}
from MSX maps integrated within a circular region of diameter $3'$.

\begin {figure}
\epsscale{1}
\plotone{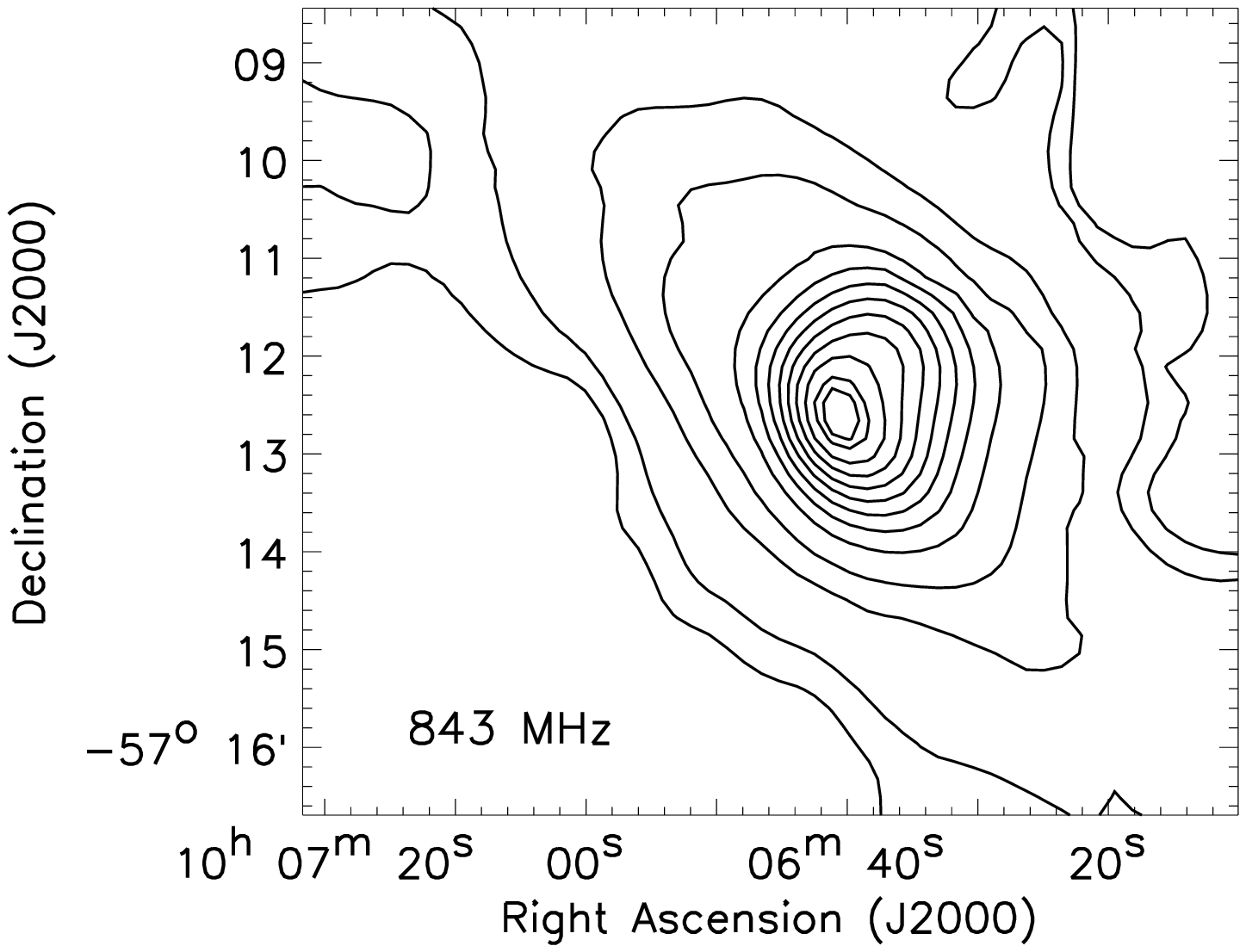}
\caption{The SUMSS radio flux density map for the region around 
\mbox{IRAS~10049-5657}
 at 843 MHz. The contour levels are at 0.1, 1, 5, 10, 20, 30, 40, 50, 60, 70, 
80, 90, 95\% of peak flux of 4.3 Jy/beam. The synthesized beam is 
$43\arcsec\times43\arcsec$.
}
\label{SUMSS_a}
\end {figure}

The SUMSS radio continuum emission from the region around 
\mbox{IRAS~10049-5657} 
at 843 MHz is shown in Fig.~\ref{SUMSS_a}. A dynamic range of $\sim700$ is 
achieved in this region (peak flux is 4.3 Jy/beam; rms noise is $\sim6$ 
mJy/beam). The radio emission peaks at 
($\alpha_{2000}$ = $10^h$ $06^m$ $41.15^s$, $\delta_{2000}$ = -57$^{\circ}$
12$\arcmin$ 40.0$\arcsec$). The integrated radio flux density is $\sim42.4$ Jy 
over 43.6 arcmin$^2$. 

\subsubsection{Stellar component}
\begin {figure}
\epsscale{1}
\plotone{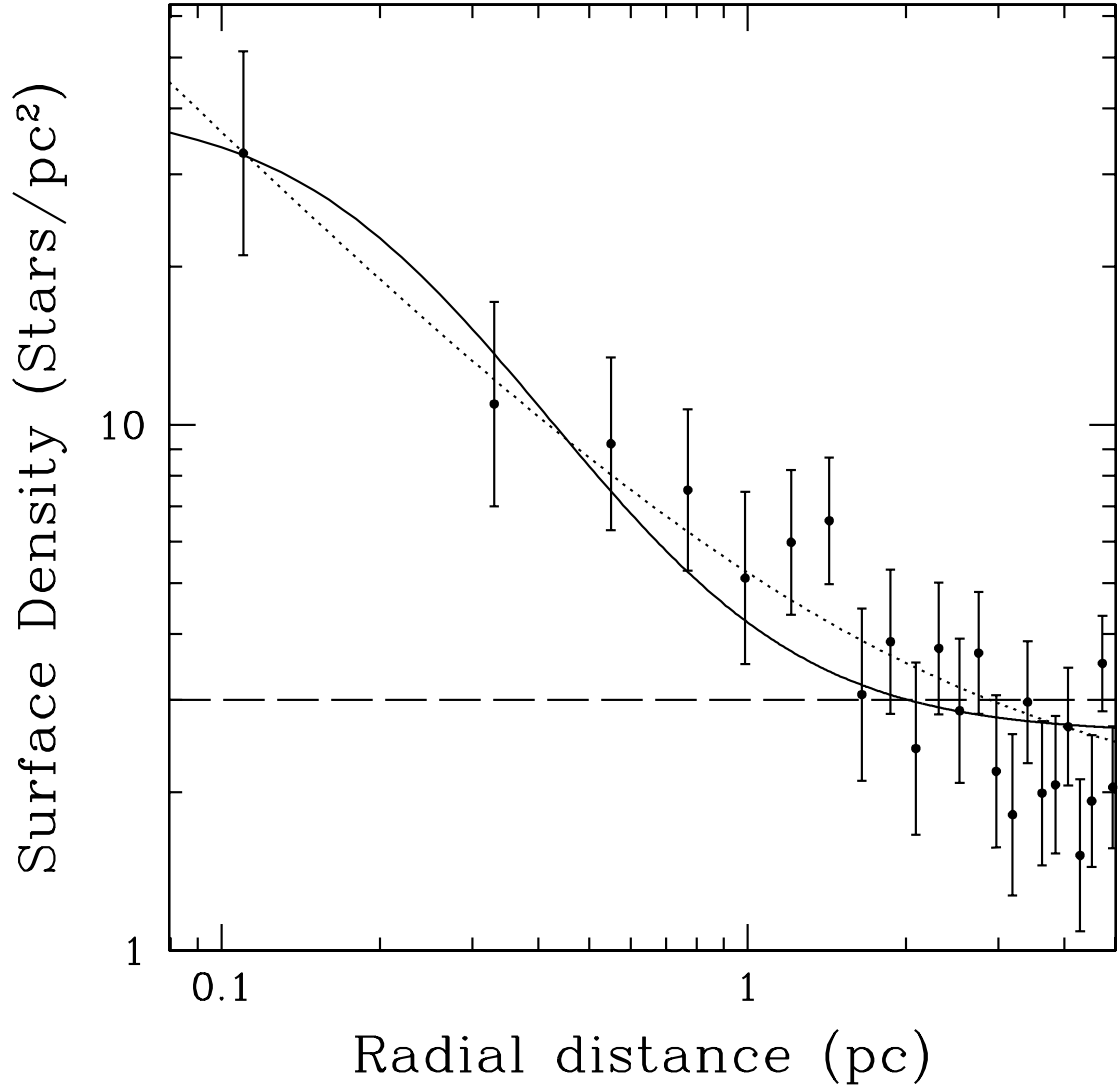}
\caption{The radial profile of the surface number density for the cluster 
associated with \mbox{IRAS~10049-5657} in log-log scale. Also plotted are the 
two fitted models - King's model (solid) and inverse radius model (dotted). 
The horizontal, dashed line corresponds to the background field 
star level, which is $\sim $3 stars pc$^{-2}$. Statistical errors 
(corresponding to the Poisson noise) are shown.
}
\label{clust_prof}
\end {figure}

The distribution of near-infrared sources (selected from 2MASS PSC) in the 
region around \mbox{IRAS~10049-5657} was investigated. A higher density of sources 
close to 
the IRAS \mbox{10049-5657} region compared to neighbouring regions implied an embedded 
cluster. We have used 2MASS sources to study the nature of embedded 
cluster. We first estimate the cluster radius. For this, we select a large 
region of radius $\sim300\arcsec$ around the IRAS peak. To account for the 
contribution from the field stars (for background determination), we 
select a control field which is 20$'$ to the east of \mbox{IRAS~10049-5657}. 
The centre of the cluster was estimated by convolving a gaussian with 
the stellar distribution and taking the point of maximum density as the 
centre. To determine the radial profile, the cluster region 
was divided into a number of concentric annuli with respect to the cluster 
centre. The surface number density of stars were obtained by counting them 
in each 7$''$ annulus and dividing 
by the annulus area. The King's model, $f(r)$, 
and the inverse radius model, $g(r)$, of the following functional forms are
fitted to the surface density radial profile.
$$f(r) = a_k + \frac{f_o}{1+(r/r_{ck})^2}$$ 
$$g(r) = a_i + \frac{a_0}{r}$$
Here, $a_k$ and $a_i$ are the fitted background constants for the King's and 
inverse radius profile, respectively. For the King's profile, $f_o$ represents
the core concentration at radius zero and $r_{ck}$ is the core radius.
 $a_0$ represents the stellar density of the inverse profile at the core.
The radial profile of the observed star density as well as the fits are shown 
in Fig.~\ref{clust_prof}. 
 A clear gradient in surface number density distribution confirms the
 existence of clustering (crowdiness) at centre of the region.
The radial profile of the cluster merges with the background field 
at $\sim 65''$, yielding the extent of the cluster to be $\sim2$ pc. 
The background level as estimated from the control field is ${\sim} 3$ 
stars~pc$^{-2}$, which agrees well with the values of $3.3\pm0.6$ and 
$\rm 2.6\pm0.2$ stars~pc$^{-2}$ yielded by the inverse radius and King's 
profile fitting, respectively.  Note that these estimates of the
stellar density towards the HII region when compared with the adjacent fields 
are unlikely to be affected by extinction (i.e. background stars) since the 
cluster is fairly distant. 

\begin {figure}
\epsscale{1}
\plottwo{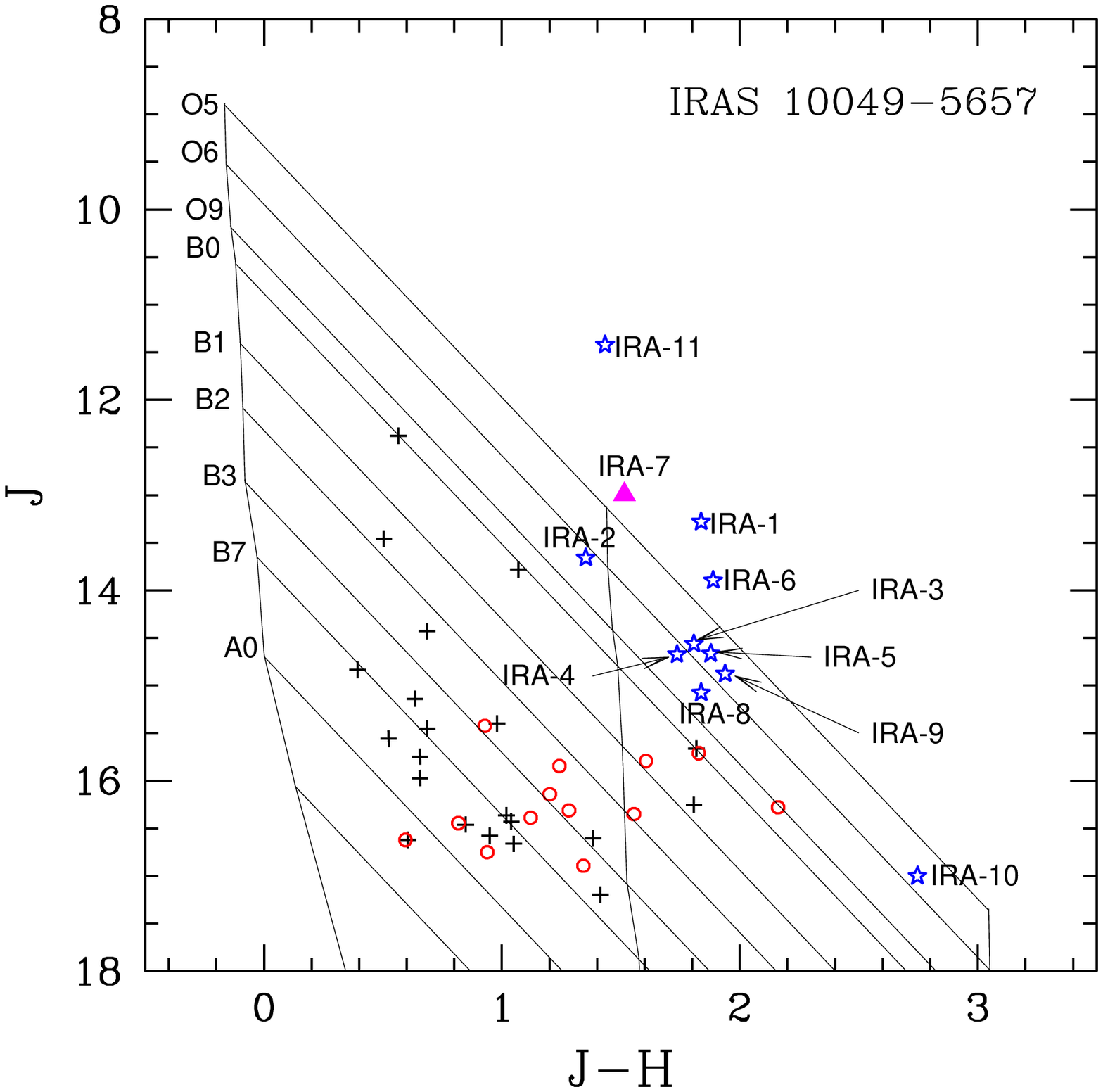}{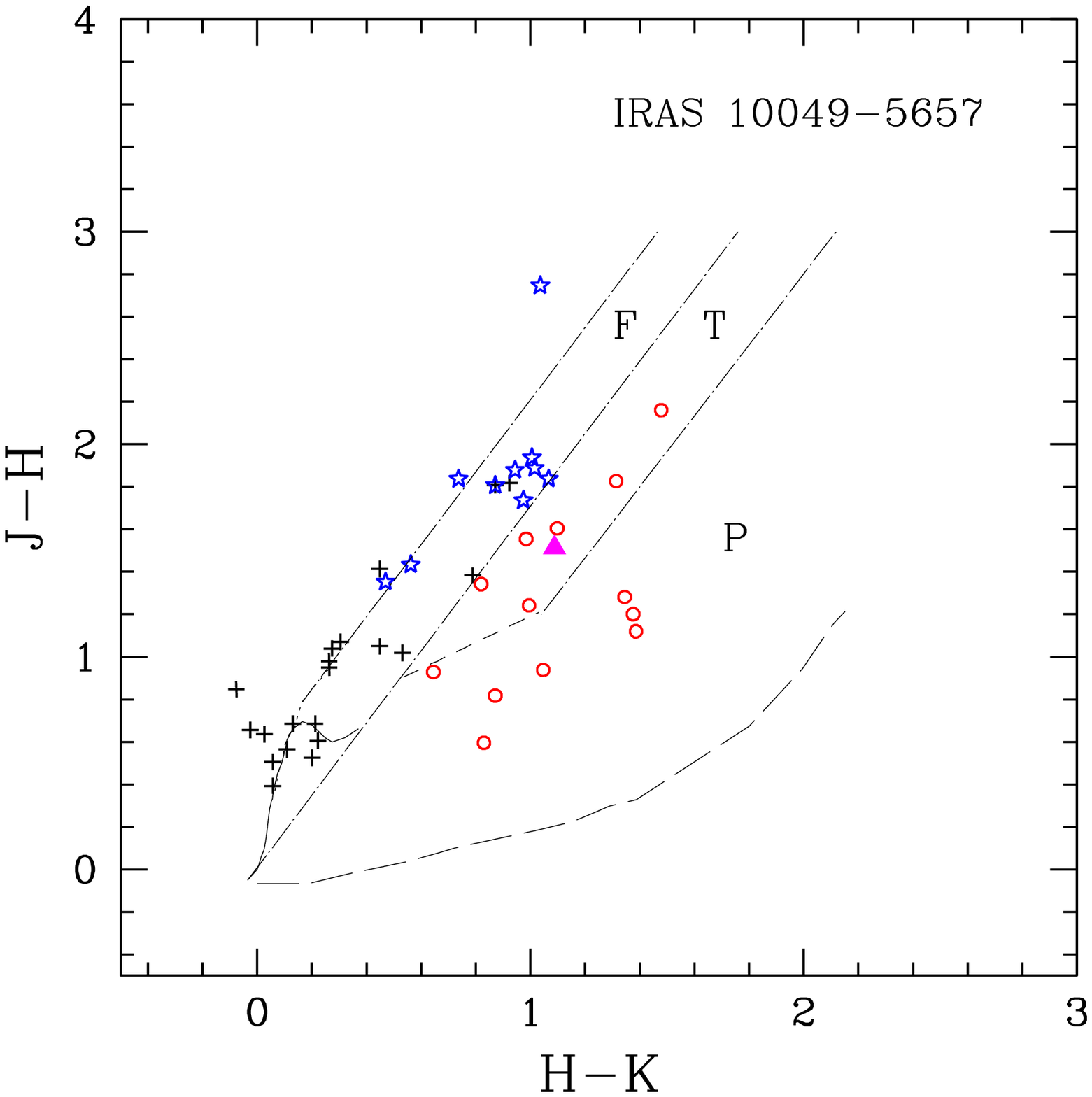}
\caption{ Colour-magnitude (left) and colour-colour diagram (right) for sources
detected in all the three 2MASS bands for the region around 
\mbox{IRAS~10049-5657}. 
In the colour-magnitude diagram, the nearly vertical solid lines from left to 
right represent the zero age main sequence (ZAMS) curves reddened by A$_V$ = 0,
 15 and 30 magnitudes, respectively. The slanting lines trace the reddening 
vectors of these ZAMS stars. In the colour-colour diagram, the locii of the 
main sequence and giant branches are shown by the solid and dotted line, 
respectively. 
The short-dash line respresents the locus of classical T Tauri stars. 
The two parallel dot-dash straight lines follow the reddening vectors 
drawn from the base of curves of giants, main-sequence stars and T-Tauri stars. 
The long-dash line represents the locus of 
Herbig Ae/Be stars. The colour-colour diagram is classified into three 
different regions, namely `F', `P' and 'T' (see text for details). The 
asterisk symbols represent sources lying above the 
ZAMS curve of spectral type O9 from the colour-magnitude diagram. The 
open circles represent sources depicting 
an infrared excess in the colour-colour diagram. The solid triangle represents 
 a source lying above the ZAMS curve of spectral type O9 as well as having an 
infrared excess i.e. lying in the `T' region. The plus symbols represent 
the other sources. 
}
\label{cmcc_a}
\end {figure}

Within a cluster radius of  $65''$ around the cluster centre 
($\alpha_{2000}$ = $10^h$ $06^m$ $38.19^s$, $\delta_{2000}$
= -57$^{\circ}$ 12$\arcmin$ 25.2$\arcsec$), 89 sources were found.
Of these, 45 were detected in all the three JHK$_s$ bands of 2MASS.  
Forty-four sources have been detected in either H and K$_s$ bands 
or only K$_s$ band. We have discussed the nature of stellar populations in 
this region using the CM (J-H vs J) and the CC (J-H vs H-K) diagrams 
which are shown in Fig.~\ref{cmcc_a}. We have 
assumed extinction values of A$_J$/A$_V$ = 0.282, A$_H$/A$_V$ = 0.175 and 
A$_{K_s}$/A$_V$ = 0.112 from Rieke \& Lebofsky (1985). All the 2MASS 
magnitudes as well as the curves are in the Bessel \& Brett (1988) system. In 
the CM diagram, the nearly vertical solid lines from left to right (with 
increasing J-H) represent the ZAMS curves (for a distance of 6.3 kpc) reddened 
by A$_V$ = 0, 15 and 30 mag, respectively. The slanting lines joining 
them trace the 
reddening vectors of these ZAMS stars. In the CC diagram, the locii 
of the main sequence and giant branches are shown by the solid and dotted 
lines, respectively. The short-dash line represents the locus of 
classical T-Tauri stars (Meyer et al., 1997). The parallel dot-dash 
straight lines follow their reddening vectors. The long-dash line represents 
the locus of Herbig Ae/Be stars (Lada and Adams, 1992).
We classify the CC diagram into three regions, as shown in the figure (see 
Ojha et al., 2004). The `F' region is considered to 
be the region where field stars (main sequence stars, giants), Class III 
and Class II objects with small infrared excess are 
located. 
The `T' sources are located redward of region `F' but blueward 
of the reddening line projected from the red end of the T-Tauri locus of Meyer 
et al. (1997). These sources may be considered to be mostly Class II with 
large NIR excess (Classical T Tauri like-stars) and/or extremely reddened early type ZAMS
stars, having excess emission in K-band (e.g., Blum, Damineli \& Conti, 2001 and 
references therein).
The `P' region is where the protostar-like Class I objects and Herbig Ae/Be 
stars are mostly located. In Fig.~\ref{cmcc_a} (left) CM diagram, the
10 sources lying above the ZAMS curve of spectral type O9 are shown as asterisk 
symbols while 13 sources with infrared excess are shown as open circles. 
It is likely that few of these objects are foreground objects or bright 
background giants. The source lying above the ZAMS curve of O9 and lying in the
 'T' region is shown 
as a solid triangle. The remaining sources are represented by plus-symbols. It 
is important to note that the CM and CC diagrams are useful tools for 
estimating the approximate nature of the stellar populations within the 
cluster in a statistical sense.

\subsection{IRAS~10031-5632} 

\subsubsection{Interstellar medium}
\begin {figure}
\epsscale{1}
\plottwo{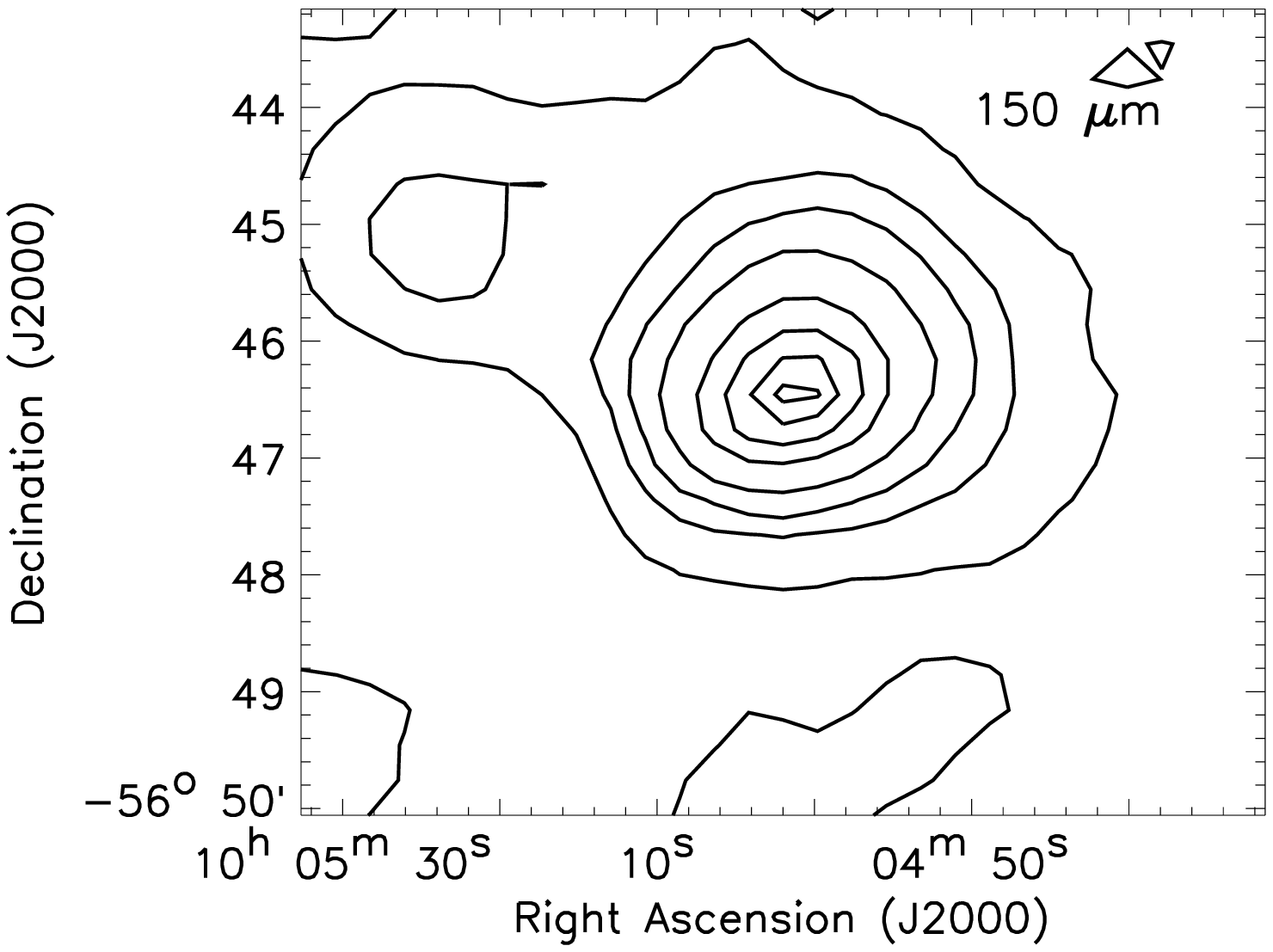}{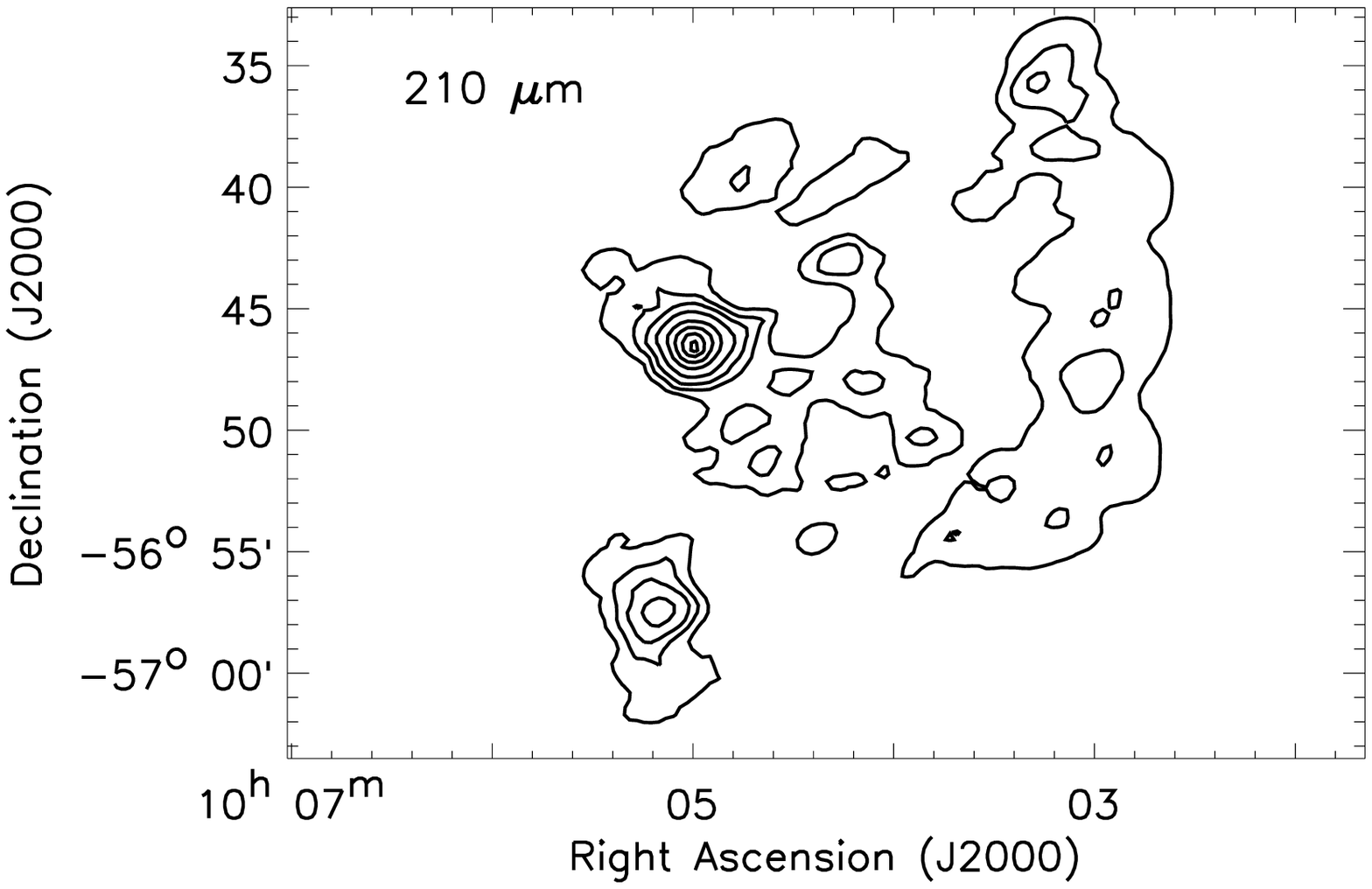}
\caption{The intensity map for the region around \mbox{IRAS~10031-5632} at 150 $\mu$m
(left) and 210 $\mu$m (right). While the 150 $\mu$m map shows emission 
from \mbox{IRAS~10049-5657} region, the 210 $\mu$m map shows the complete 
region scanned by the far-infrared telescope. Contour levels are at 1, 5, 10, 
20, 40, 60, 80, 
95\% of peak intensity of 712 Jy/sq arc min (left) and 220 Jy/sq arc min 
(right).
}
\label{FIRmap_b}
\end {figure}

\begin {figure}
\plottwo{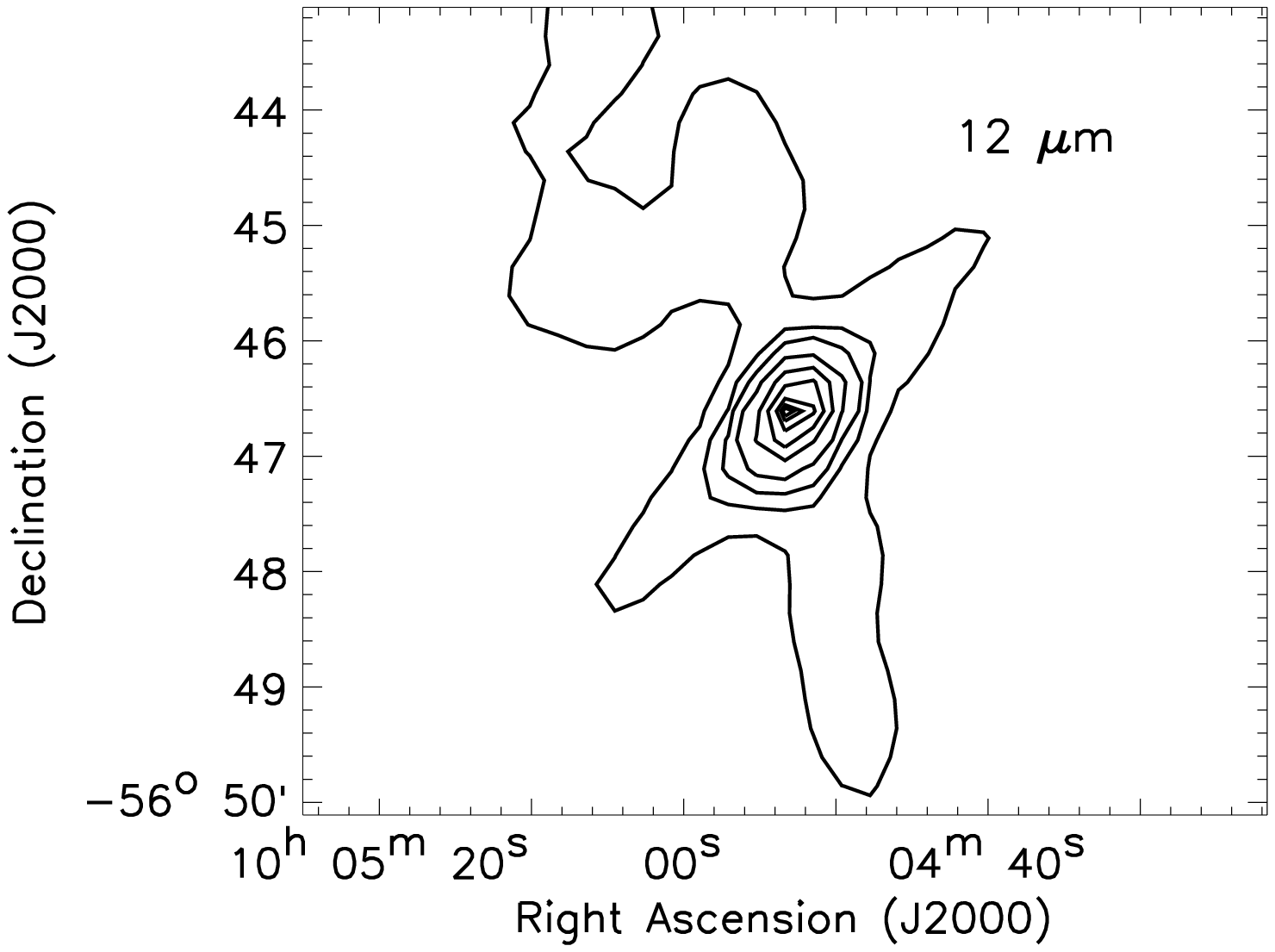}{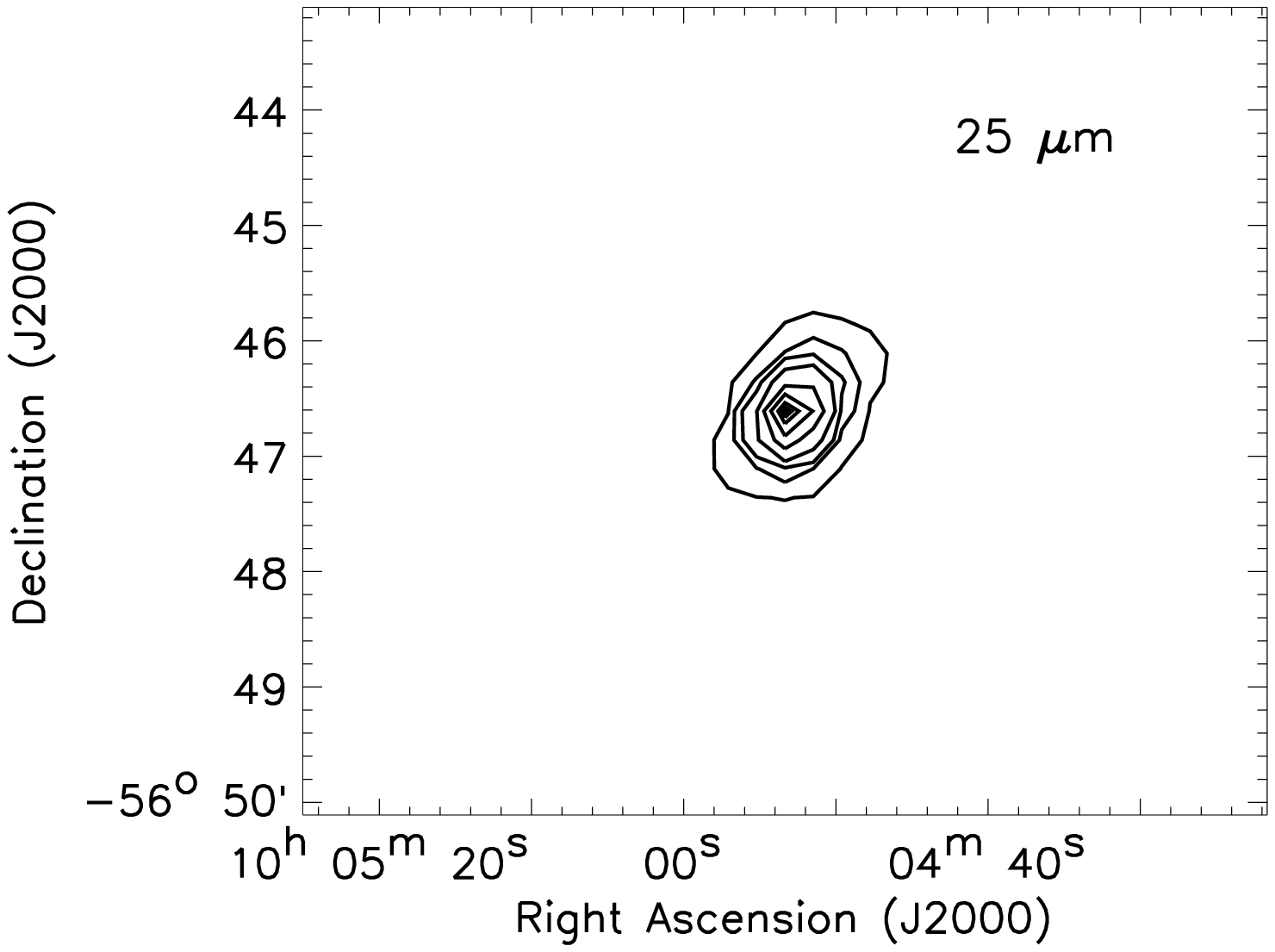}
\plottwo{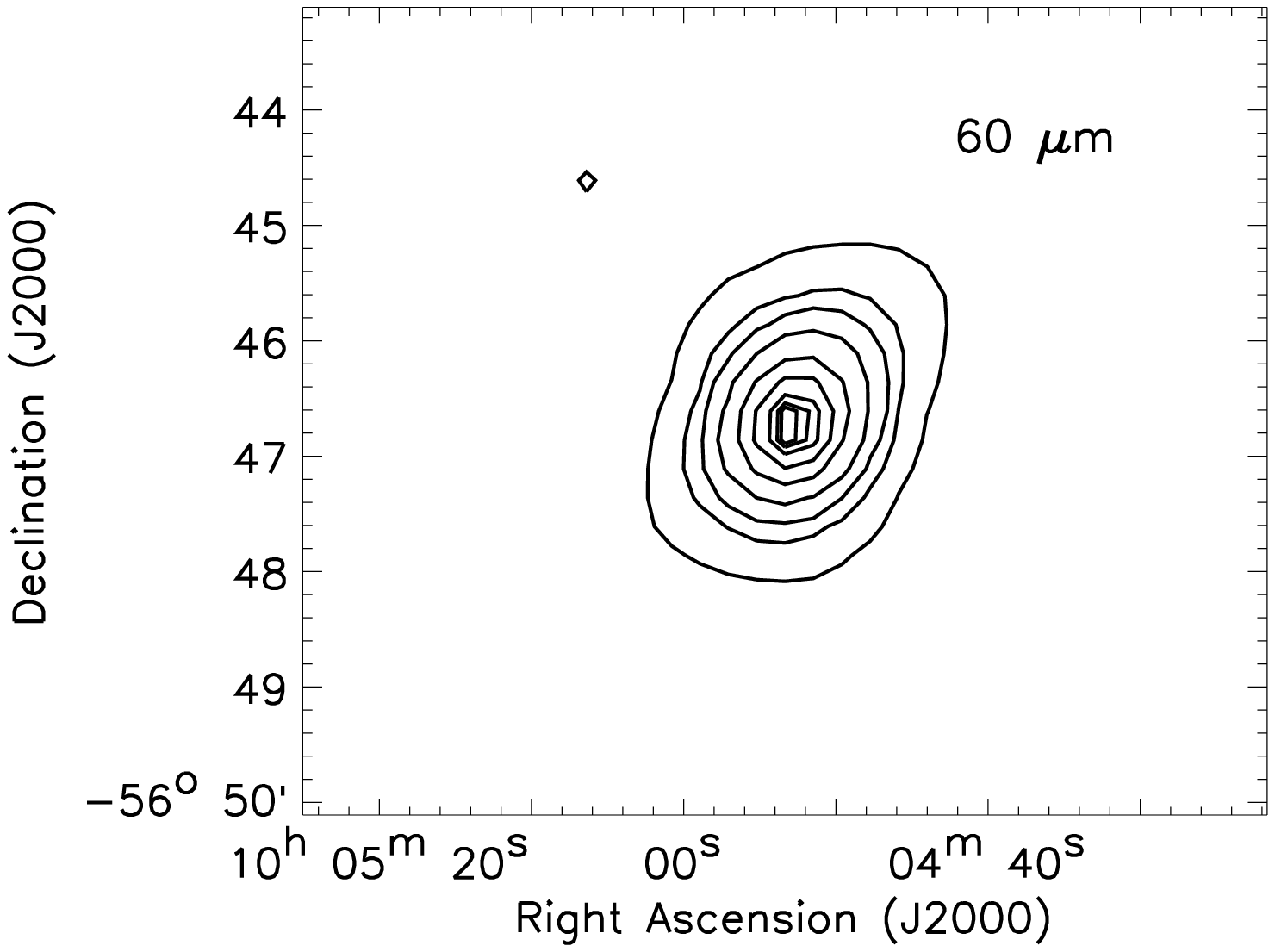}{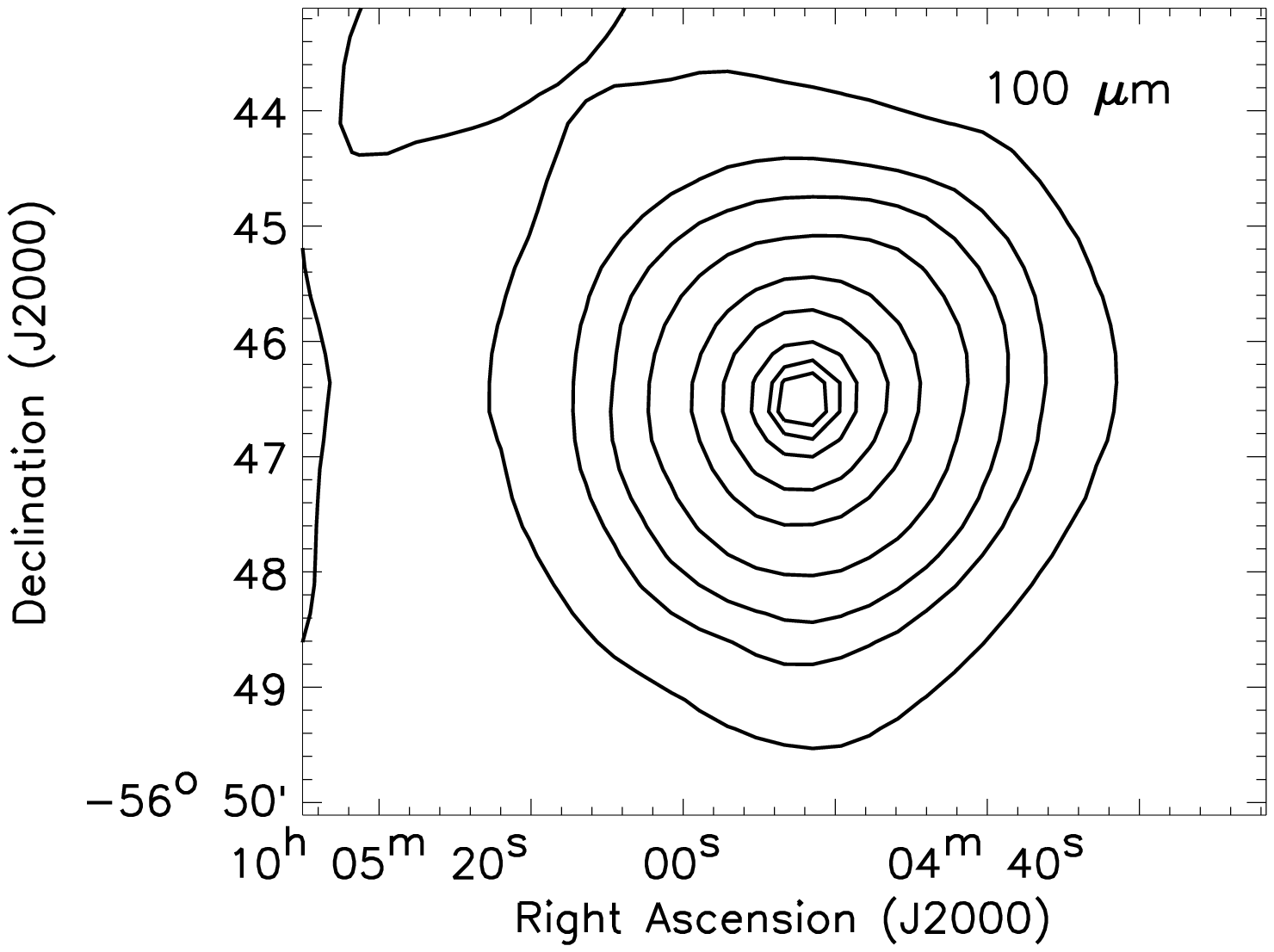}
\caption{The IRAS-HIRES intensity maps for the region covering 
\mbox{IRAS~10031-5632}
at 12 $\micron$ (top left), 25 $\micron$ (top right), 60 $\micron$ (bottom
left) and 100 $\micron$ (bottom right). The contours in the emission maps are 
at 1, 5, 20, 40, 60, 80, 90 and 95 \% of the peak value of 41.8 Jy 
arcmin$^{-2}$, 774 Jy arcmin$^{-2}$, 1370 Jy arcmin$^{-2}$ and 486 Jy 
arcmin$^{-2}$ at 12 and 25, 60 and 100 $\mu$m respectively. 
}
\label{HIRES_b}
\end {figure}

The intensity maps of the region around \mbox{IRAS~10031-5632} in the 150 and 
210 $\mu$m FIR bands are shown in Fig.~\ref{FIRmap_b}. The emission at
 210 $\mu$m from the entire mapped region ($25'\times18'$) is shown in 
Fig.~\ref{FIRmap_b} (right). The isophotes in both the maps have been 
displayed upto 1\% level of the peak intensities. 
A detailed investigation of these maps reveals that \mbox{IRAS~10031-5632} is 
unresolved at 150 $\mu$m but is extended at 210 $\mu$m map. 
At fainter levels, we observe large-scale extended emission, 
particularly towards the south-east.
The HIRES-processed IRAS maps for the corresponding region 
around \mbox{IRAS~10031-5632} are shown in Fig.~\ref{HIRES_b}. The angular 
resolution achieved in these maps are $0\arcmin.9\times0\arcmin.5$ at 12 and 
25 \micron, $1\arcmin.4\times1\arcmin.1$ at 60 $\micron$, 
$2\arcmin.1\times1\arcmin.9$ at 100 $\micron$. The peak position and flux 
density details of \mbox{IRAS~10031-5632} are given in Table~\ref{fluxes}. 
The emission in the mid infrared bands of MSX has been used to model the 
the peak UIB emission which is 
$6.5\times10^{-5}$ W m$^{-2}$ Sr$^{-1}$. 

\begin {figure}
\epsscale{1}
\plotone{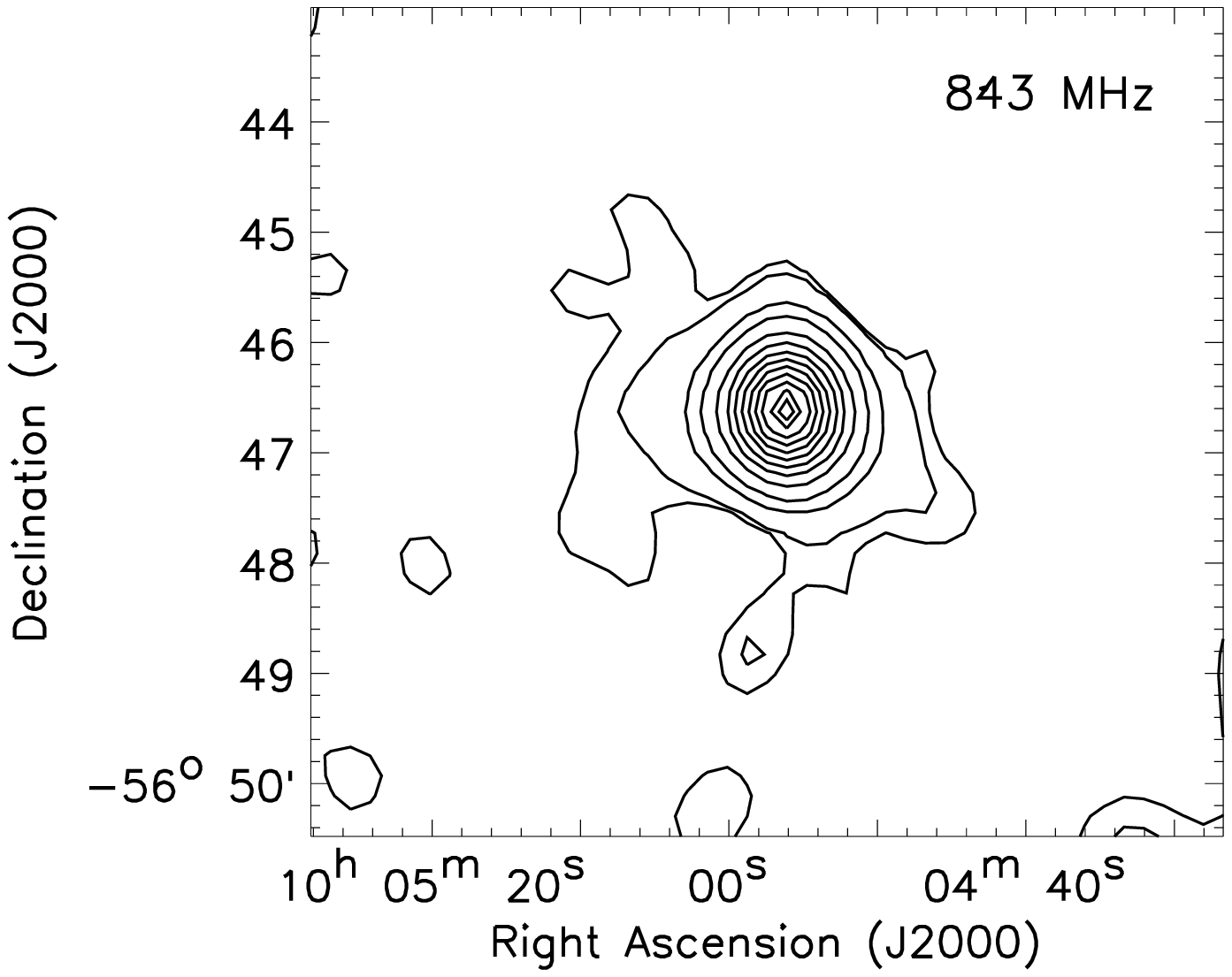}
\caption{The SUMSS radio flux density map for the region around 
\mbox{IRAS~10031-5632}
at 843 MHz. The contour levels are at 0.5, 1, 5, 
10, 20, 30, 40, 50, 60, 70, 80, 90, 95\% of peak flux of 0.65 Jy/beam. The 
beam is $43\arcsec\times43\arcsec$.
}
\label{SUMSS_b}
\end {figure}

The SUMSS radio emission in region around \mbox{IRAS~10031-5632} at 843 MHz is 
shown in Fig.~\ref{SUMSS_b}. The dynamic range of the radio map is $\sim300$ 
(peak flux is 652 mJy/beam; rms noise is $\sim2$ mJy/beam). The peak of radio 
emission is at $\alpha_{2000}$ = $10^h$ $04^m$ $56.10^s$, $\delta_{2000}$ = -56$^{\circ}$ 46$\arcmin$ 38.0$\arcsec$. The integrated flux density is $1.1$ Jy 
over 4.9 arcmin$^2$.

\subsubsection{Stellar component}
\begin {figure}
\epsscale{1}
\plottwo{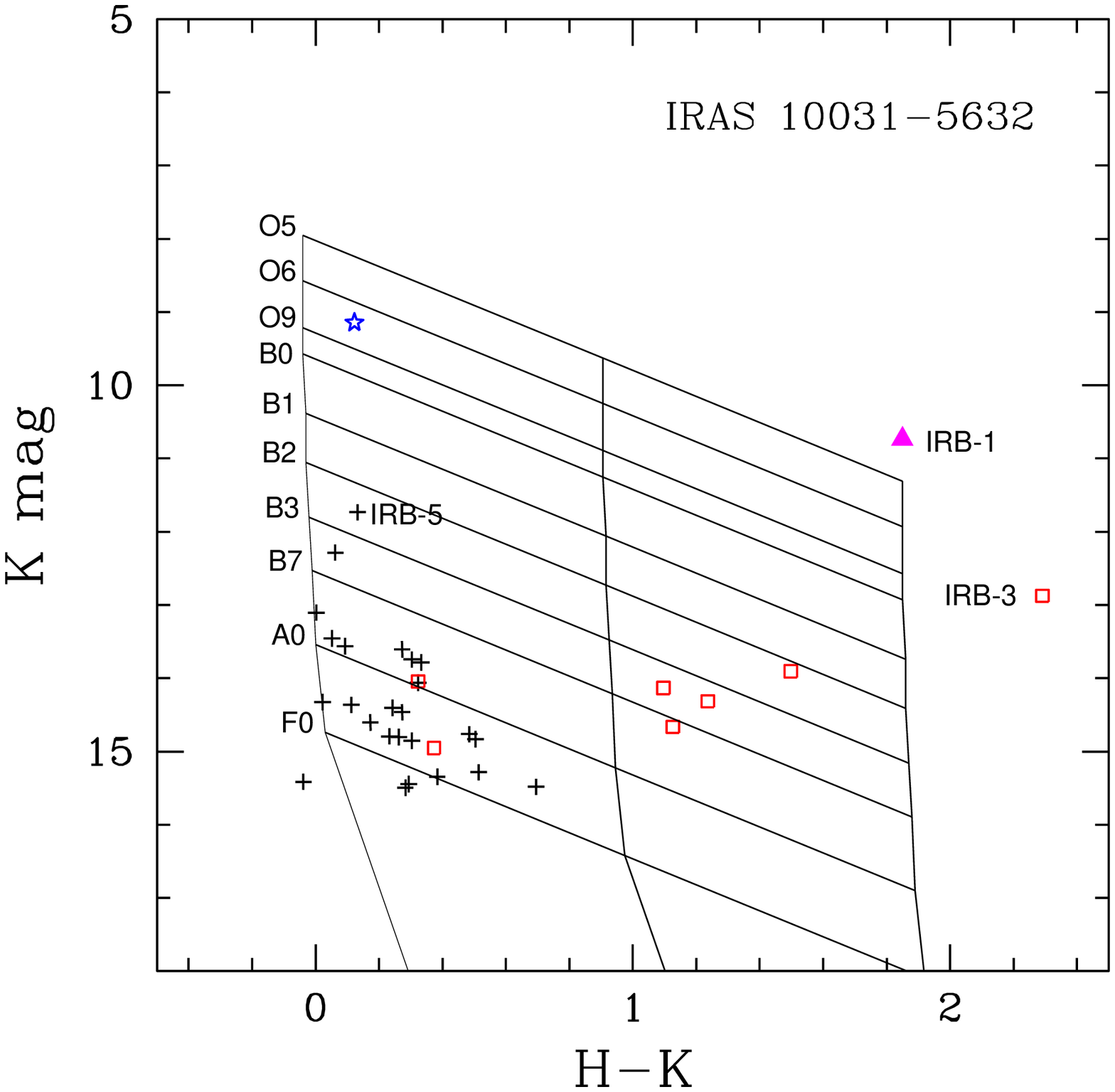}{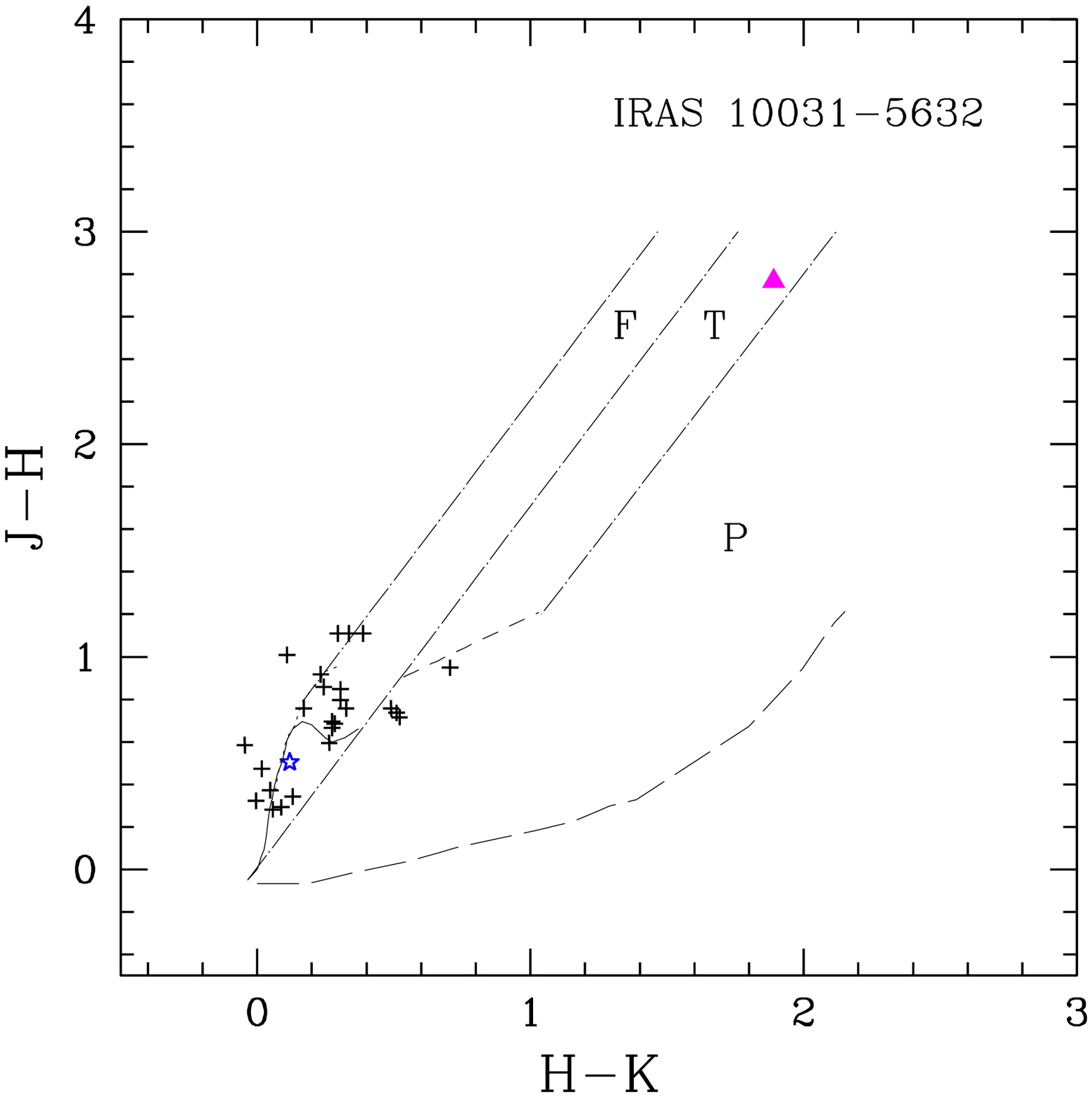}
\caption{ Colour-magnitude (left) and colour-colour diagram (right) for sources
detected in all the three 2MASS bands for the region around 
\mbox{IRAS~10031-5632}.
In the colour-magnitude diagram, the nearly vertical solid lines from left to
right represent the zero age main sequence (ZAMS) curves reddened by A$_V$ = 0,
 15 and 30 magnitudes, respectively. The slanting lines trace the reddening
vectors of these ZAMS stars. In the colour-colour diagram, the locii of the
main sequence and giant branches are shown by the solid and dotted line,
respectively. 
The short-dash line respresents the locus of classical T Tauri stars. 
The three parallel dot-dash straight lines follow the reddening vectors 
drawn from the base of curves of giants, main-sequence stars and T-Tauri stars.
The long-dash line represents the locus of
Herbig Ae/Be stars. The colour-colour diagram is classified into three
different regions, namely `F', `P' and 'T' (see text for details). The 
asterisk symbols represent sources lying above the
ZAMS curve of spectral type O9 from the colour-magnitude diagram. The 
open circles represent sources depicting
an infrared excess in the colour-colour diagram. The solid triangle represents
 a source lying above the ZAMS curve of spectral type O9 as well as having an
infrared excess. The plus symbols represent 
the other sources. The open squares in the CM diagram represent the 2MASS 
sources detected in only H and K$_s$ bands.}
\label{cmcc_b}
\end {figure}

We have used the 2MASS sources in the vicinity of \mbox{IRAS~10031-5632} to 
study the stellar populations here. In a 
circular region of radius $1'$ around the IRAS peak, 27 sources have been 
detected in JHK$_s$ bands while 14 sources are detected in either 
H and K$_s$ bands or only K$_s$ band. The CM (H-K vs K) and CC (J-H vs H-K) 
diagrams of the sources detected in all the three bands are shown 
in Fig.~\ref{cmcc_b}. The ZAMS curves in the CM diagram are for a distance of 
3.7 kpc reddened by A$_V$ = 0, 15 and 30 mag, respectively. As in the case of 
\mbox{IRAS~10049-5657}, all the 2MASS magnitudes as well as the curves are in 
the Bessel \& Brett (1988) system. The source lying above the ZAMS curve of 
spectral type O9 is represented by an asterisk while the filled triangle 
represents a source lying above the ZAMS curve of spectral type O9 and having 
an infrared excess. The 7 sources detected only in HK$_s$ bands are 
shown as 
open squares in the figure. The remaining sources are denoted by crosses.
An investigation into the distribution of 2MASS PSC sources in the region 
around \mbox{IRAS~10031-5632} shows a few sources grouped together near the IRAS peak.
 For this cluster,
we have been unable to determine the cluster radius owing to the low surface
density of stars (very few excess stars close to the IRAS peak).

\section{Radiative Transfer Modelling}

In an attempt to obtain a self-consistent picture of these star forming 
regions using all the available data as well as to extract important physical 
parameters, we have carried out radiative transfer modelling of 
\mbox{IRAS~10049-5657} and \mbox{IRAS~10031-5632}. The symmetric morphology of 
contours near the centrally located peak in the FIR maps supports the 
one-dimensional treatment of the radiative transfer modelling.

\subsection{Continuum emission}

Each star forming region is modelled as a spherically symmetric cloud of
gas (hydrogen) and dust, powered by a centrally embedded source (single or
cluster of ZAMS stars). The cloud is immersed in an average interstellar 
radiation field. The radiative transfer equations have been solved assuming 
a two-point boundary condition for the spherical cloud. The gas exists 
throughout the cloud, i.e. from the stellar surface to the edge of the cloud. 
The dust exists in a spherical shell with a cavity at the centre. This is 
because close to the exciting source(s), the dust grains are destroyed when 
exposed to excessive radiative heating. Two commonly used types of 
interstellar dust are explored; the DL type (Draine \& Lee, 1984) and the MMP 
type (Mezger, Mathis \& Panagia, 1982). In the spherical shell where gas and 
dust co-exist, the gas-to-dust ratio is held constant. 
Other 
details of this self-consistent scheme are given in Mookerjea \& Ghosh (1999).
The parameters explored are as follows :
the geometrical dimensions of the cloud (primarily the radius of the 
dust-free cavity; the outer diameter is guided by the observed angular 
extent and the distance), 
radial density distribution law (three power law exponents : 
n(r) $\sim$ r$^0$, r$^{-1}$, or r$^{-2}$),
radial optical depth due to dust,
the nature of embedded source/(s) (single ZAMS star or a cluster, 
consistent with the total observed luminosity), 
relative abundances of different grain types (silicate, graphite), 
and the gas-to-dust ratio.
The observational constraints include the SED due to thermal emission from the 
dust component, angular sizes at different wavelengths and radio continuum 
emission from the H II region. 
The parameters corresponding to the best fit models for the two sources
are presented in Table 2. 
While no claim is made about the uniqueness of these parameter-sets,
the following comments in support of the robustness of the results
are in order :
- a constant density radial profile provides the best fit for both regions 
and the r$^{-1}$ or r$^{-2}$
density profiles are conclusively ruled out by the observed SED;
- inner (dust-free) cavity radii lower than those derived result in
short wavelength fluxes significantly higher than those observed and 
are thus ruled out;
- radio continuum emission is quite sensitive to the nature of
 the central energy source, viz., single star versus a cluster.

\subsection{Line emission from gas}

We have also modelled the infrared nebular/ionic fine structure line emission 
from gas around \mbox{IRAS~10049-5657} and \mbox{IRAS~10031-5632},
using a sophisticated scheme 
which uses more details of the interstellar gas component.
This scheme developed by Mookerjea \& Ghosh (1999), uses the  
photoionisation code CLOUDY (Ferland, 1996),
which solves for statistical and thermal equilibrium by balancing
various ionization-neutralization processes as well as heating-cooling
processes.
The scheme additionally includes : 
 (a) the exact structure of the cloud / H II region 
(viz., central dust-free cavity) and 
(b) absorption effects of the dust component on the emergent line intensities. 
Typical H II region abundance of the 
gas component has been taken into consideration. 
The overall structure of the cloud is defined by the
parameters that provide the best fit to the continuum spectral energy
distribution (see subsection 4.1).
The first part of the calculations involve
the pure gas inner shell and the emerging 
spectrum comprises of continuum as well as line emission. This emergent
continuum from the inner shell 
provides the inner surface boundary condition for
the second shell comprising
of gas and dust. The line emission from the inner shell is transported 
outwards through extinction by the dust column in the second shell.
Similarly, line emission originating from the second shell
is transported considering the absorption effects of the dust
grains lying between the emission zone and the outer surface of
the second shell.
The finally predicted emergent line luminosity includes both these components.
A total of 27 spectral
lines in the wavelength range $2.5-200$ $\mu$m have been considered. 
To compare the predictions of the model with the spectral lines detected 
in the IRAS-LRS spectra of \mbox{IRAS~10049-5657} and \mbox{IRAS~10031-5632}, 
we convolve the 
model predicted spectrum with IRAS-LRS instrument profile in the wavelength 
range 8-22 $\mu$m. However, to 
show the complete spectrum for the wavelength range 2.5-200 $\mu$m, the
spectral lines have been convolved with ISO-SWS and LWS typical spectral
resolutions of 1000 for $2.5\le\lambda<12$ $\mu$m, 20000 for $12\le\lambda<45$ 
$\mu$m, 8100 for $45\le\lambda<75$ $\mu$m and 6800 for $75\le\lambda<200$ 
$\mu$m while predicting the expected emergent spectrum. 

\subsection{IRAS~10049-5657}

\begin {figure}
\epsscale{1}
\plottwo{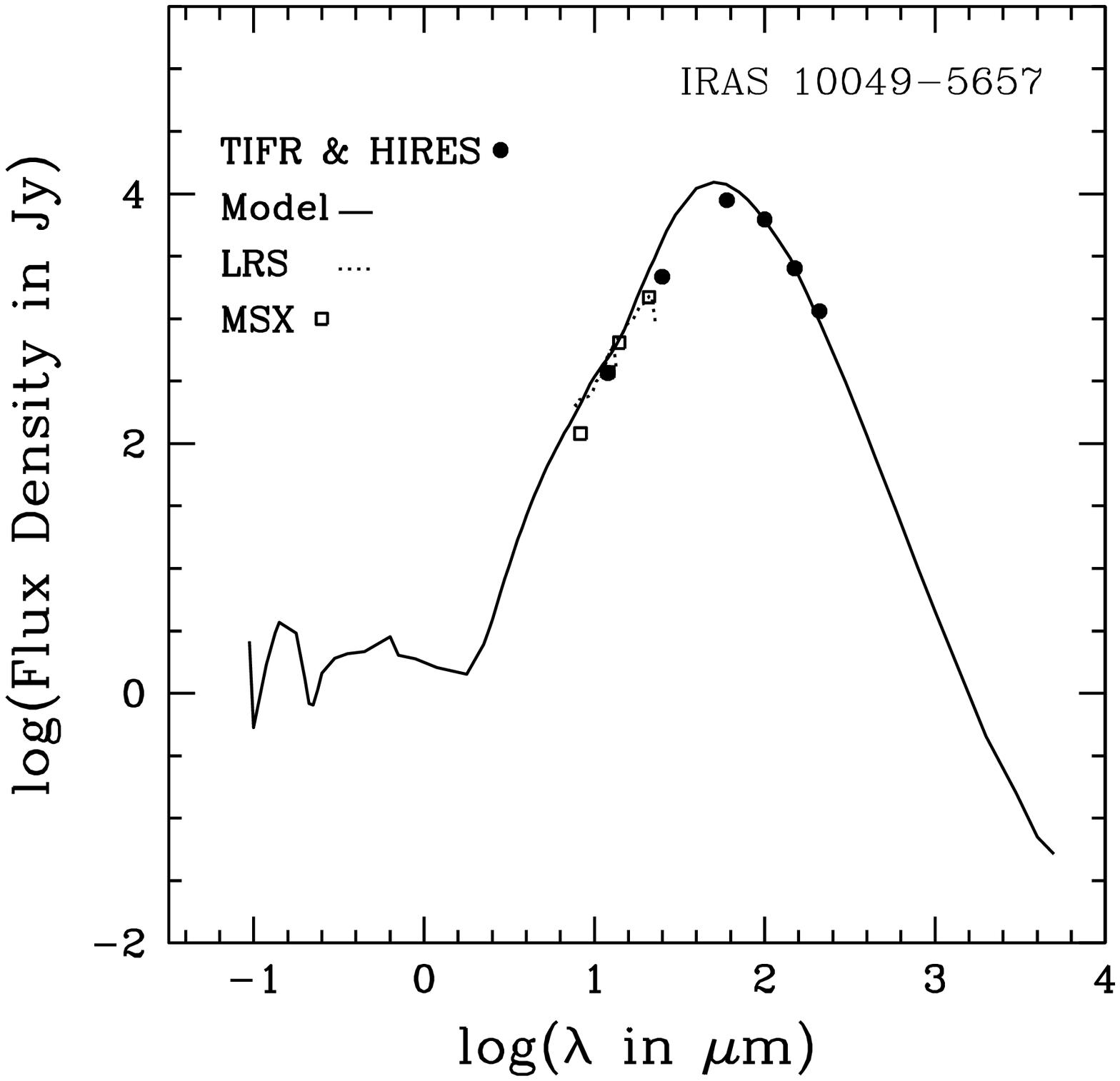}{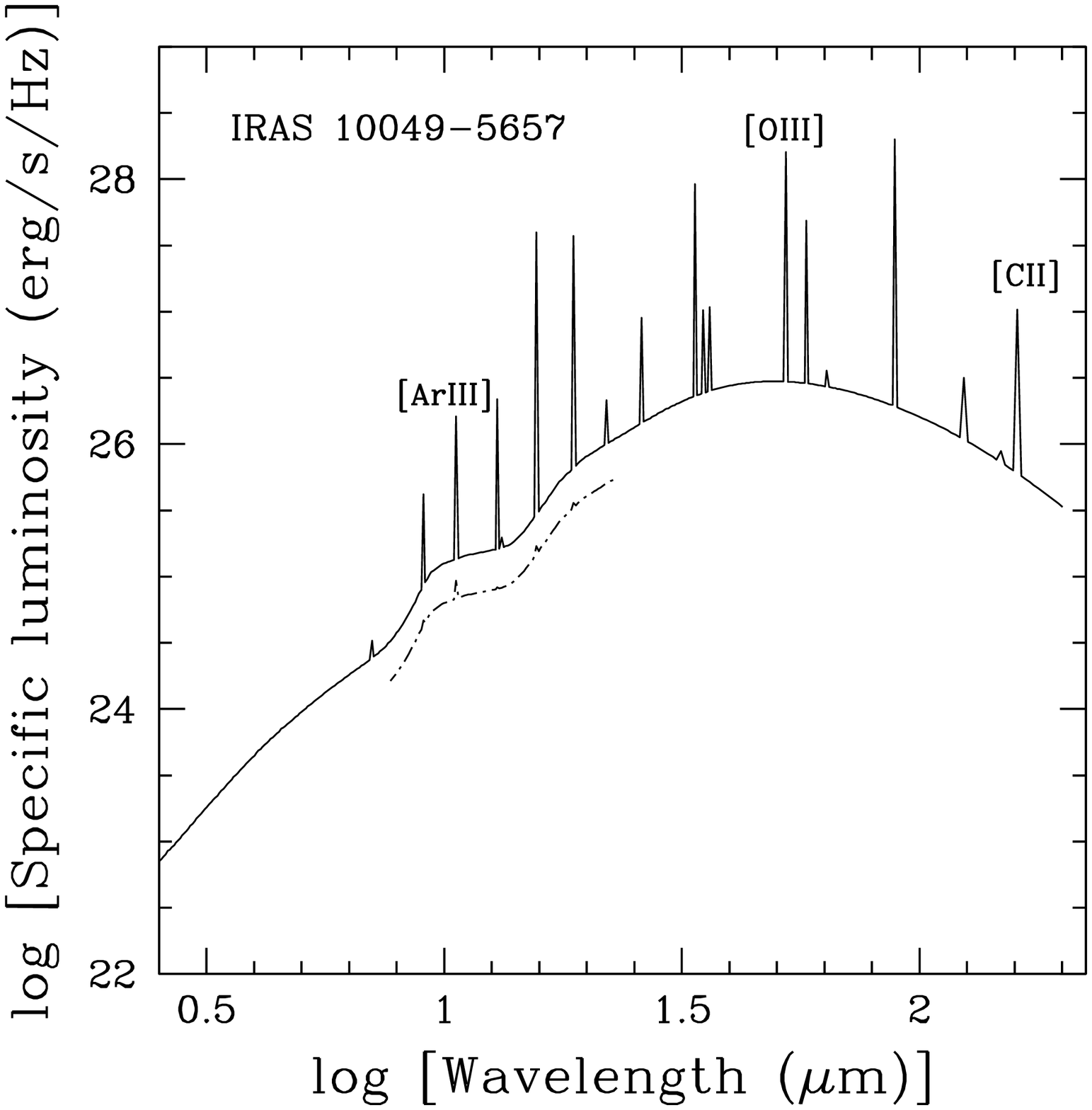}
\caption{(Left) Comparison of the spectral energy distribution from
observations and the best fit radiative transfer model of 
\mbox{IRAS~10049-5657} for continuum emission. The filled circles represent 
the TIFR and IRAS-HIRES data. The dashed lines are
the IRAS-LRS spectra and the open squares represent the MSX data.
The solid line presents the predicted emergent spectrum from our radiative
transfer model. See text and Table~\ref{radtran_parm} for details of model 
parameters. (Right) Emergent spectrum predicted by the scheme for line emission
 modelling for \mbox{IRAS~10049-5657}. The solid line represents the model 
convolved with the ISO spectral resolutions while the dot-dashed line 
(shifted along Y-axis for better viewing) represents convolution with the 
LRS-instrument profile. 
}
\label{radtran_1004}
\end {figure}

The SED for \mbox{IRAS~10049-5657} is constructed using the flux densities in 
the two TIFR bands,  the four IRAS bands (from HIRES maps), the IRAS-LRS data
and the four MSX bands. The flux densities used in the SED are the fluxes 
integrated over a circular region of diameter $3'$ around the peak. The total 
luminosity is $1\times10^6$ L$_{\odot}$ for a distance of 6.3 kpc. 
Fig.~\ref{radtran_1004} (left) shows the observed SED and the predicted 
spectrum from the best fit model. This best fit model implies 
a uniform dust and gas density distribution with
the embedded energy source as a single ZAMS star of type O5-O4 (see Table 2).
The inner cloud 
dust radius is 0.008 pc while the outer cloud radius is 4.2 pc. The optical 
depth at 100 $\mu$m is 0.002. The radius of the ionised gas from the model is 
2.8 pc. The observed angular sizes are explained by this model. The measured 
radio flux density, $\sim25$ Jy at 843 MHz is obtained by integrating within a 
circular region of 2.8 pc around the radio peak. This flux is quite high and 
cannot be explained by the model as the predicted radio flux from the model is 
4 Jy for gas-to-dust ratio of 100:1 by mass. The dust composition for Si:Gr is 
11:89 for the DL type of dust. The total dust mass is determined to be 12 
M$_\odot$.

The modelling of line emission for \mbox{IRAS~10049-5657} has been carried out 
using 
the physical sizes obtained from the modelling of continuum emission. In all, 
21 nebular/ionic lines satisfy our detectability criterion 
(power in the line be at least 1\% of the power in the neighbouring 
continuum). The wavelengths and luminosities of these lines are presented in
Table~\ref{model_linea}. The modelled spectrum, including the lines 
from the 10 elements considered as well as the continuum predicted by this 
model is shown in Fig.~\ref{radtran_1004} (right). 

We have looked for signatures of the ionic lines generally found in HII regions
 in the IRAS-LRS spectrum using a simple method. This method assumes the 
line signal to be present within data points, D${}_{line}$,
corresponding to one resolution element of LRS, centered at the expected
line position. The local spectral baseline is estimated by interpolating
(power law) data points in one resolution element each on both the
{\it red} \& {\it blue} sides of D${}_{line}$. The estimated baseline is
then subtracted from D${}_{line}$ to obtain the line emission.
For \mbox{IRAS~10049-5657}, the `brighter' lines: [Ne~II] at 12.8 $\mu$m, [Ne~III] at 
15.5 $\mu$m, [S~III] at 18.7 $\mu$m and [Ar~III] at 21.8 $\mu$m are detected. 
The luminosities of these lines as well as the ratios of line-to-continuum are 
listed in Table~\ref{model_linea}. 

We compare the luminosities as well as the ratios of line-to-continuum of the 
predicted lines from the model (convolved with IRAS-LRS instrument 
profile) with respect to those observed. 
The line-to-continuum ratios for the `LRS-convolved' model are listed in 
Table~\ref{model_linea}. We find that the ratios of luminosities 
(model/observations) for the lines [Ne~II], [Ne~III] and [S~III] 
agree within a factor of 4. For the [Ar~III] line, we find that the 
observed value is $\sim13$ times larger than the modelled value. 
However, it is to be noted 
that [Ar~III] line is close to the edge of the wavelength range of the 
spectrometer and could have instrumental uncertainities. The lines [Ar~III] at 
9.0 $\mu$m and [S~IV] at 10.5 $\mu$m are barely detected although the model 
predicts them to be bright.

\subsection{IRAS~10031-5632}

\begin {figure}
\epsscale{1}
\plottwo{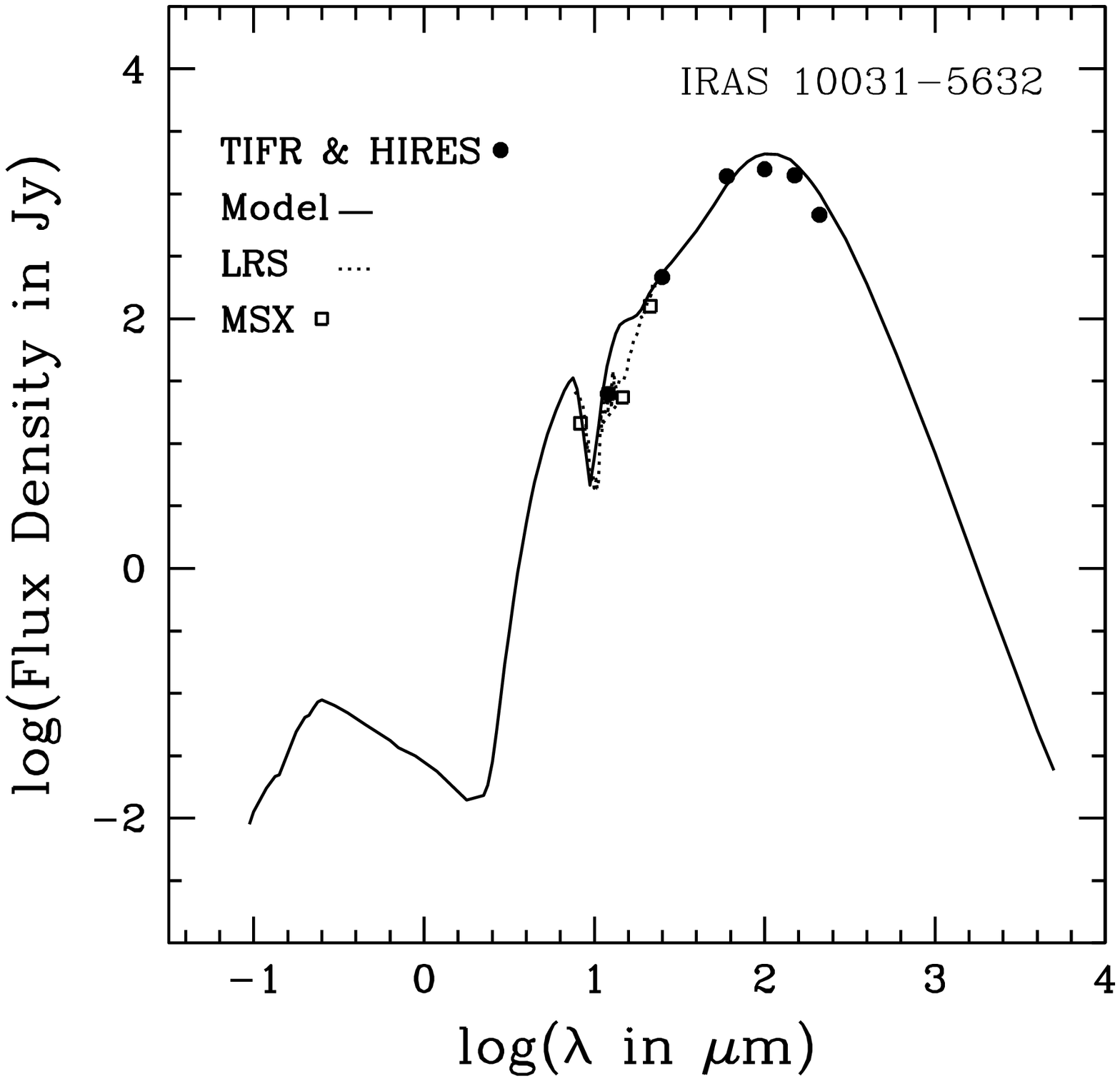}{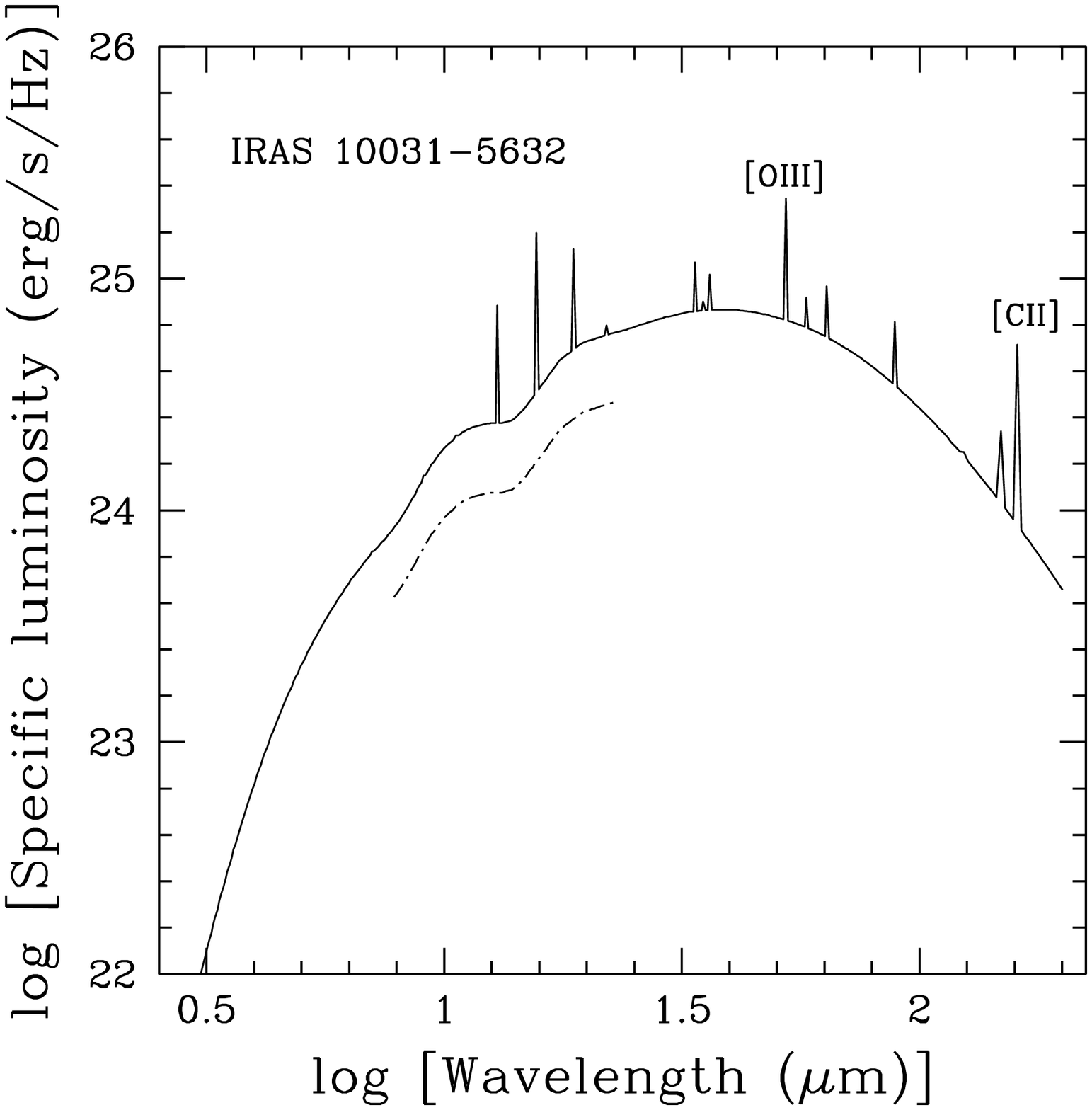}
\caption{(Left) Comparison of the spectral energy distribution from
observations and the best fit radiative transfer model of
\mbox{IRAS~10031-5632} for continuum emission. The filled circles represent
the TIFR and IRAS-HIRES data. The dashed lines are
the IRAS-LRS spectra and the open squares represent the MSX data.
The solid line presents the predicted emergent spectrum from our radiative
transfer model. See text and Table~\ref{radtran_parm} for details of model 
parameters. (Right) Emergent spectrum predicted by the scheme for line 
emission  modelling for \mbox{IRAS~10031-5632}. The solid line represents the 
model convolved with the ISO spectral resolutions while the dot-dashed line 
(shifted along Y-axis for better viewing) represents convolution with the 
LRS-instrument profile.
}
\label{radtran_1003}
\end {figure}

The SED for \mbox{IRAS~10031-5632} has been constructed using flux densities 
(integrated over a circular region of $3'$ diameter centred on the peak) from 
the MSX, IRAS-HIRES as well as the TIFR maps. The IRAS-LRS spectrum has also 
been used in the construction of the SED. The spectrum shows a strong silicate 
feature. For \mbox{IRAS~10031-5632}, the 
total luminosity is $4.2\times10^4$ L$_{\odot}$ (d$\sim3.7$ kpc). We have used 
a single ZAMS star of spectral type O9 as the centrally exciting source. The 
best fit radiative transfer model along with the observed SED is shown in 
Fig.~\ref{radtran_1003} (left), and the parameters of the best fit model are 
given in Table~\ref{radtran_parm}. This model implies a uniform 
density distribution of gas and dust. The optical depth at 100 $\mu$m (from the
model) is 0.07. The ratio of silicates and graphite dust grains is estimated 
to be 62:38. The radio flux predicted by the model at 843 MHz is 0.07 Jy for a 
gas-to-dust ratio of 100 by mass and radius of ionised region is 0.03 pc. This 
is comparable to the measured radio 
flux density of 0.04 Jy at 843 MHz (obtained by integrating the flux within 
a circular region of radius 0.03 pc centred on the radio peak). 
A dust mass of 24 $M_{\odot}$ is obtained from the model.

For \mbox{IRAS~10031-5632}, 17 nebular/ionic lines satisfy the detectability 
criterion. The luminosities of the lines as well as the ratio of luminosities 
of each line with respect to the continuum  (for the model convolved with 
the LRS spectral resolution) are also listed in Table~\ref{model_lineb}. 
Figure~\ref{radtran_1003} (right) shows the emerging spectrum 
from the model.  
In addition, following the method in Sect. 4.3, we have searched for lines 
in the LRS spectrum of \mbox{IRAS~10031-5632}. The [Ne~II] line at 12.8 $\mu$m is 
clearly detected with a luminosity of 30 L$_{\odot}$. Among the other lines 
in the range of LRS, we find that [Ar~III] at 9.0 $\mu$m and [S~III] at 18.7 
$\mu$m also show detections. A comparison of the luminosites of the 
detected lines 
with the predicted values shows that the observed luminosities are much higher 
than the modelled ones. Table~\ref{model_lineb} also lists the ratio of 
line-to-continuum for these lines. A comparison of these ratios shows that 
the model ratios are a factor of $\sim100$ lower than those observed. 
This can be attributed to the elevated level of the continuum from the 
model. A comparison of the model SED with the 
observed one in Fig.~\ref{radtran_1003} (left) shows that the radiative 
transfer model in the mid infrared overestimates the observed LRS spectrum. 
Much larger discrepancies between the observed and predicted line emissions 
for \mbox{IRAS~10031-5632}, may be understood as follows.
Our scheme for predicting the line emission utilizes the   
description of the cloud, which is obtained from modelling the continuum 
emission from the dust component distributed throughout
the cloud. In contrast, emission of the specific ionic lines under discussion,
viz., [S III], [Ne II] and [Ar III] are expected to originate from the
innermost part of the H II region (due to high ionization potentials).
In addition, the lower excitation of the central source (ZAMS O9) for
this source, makes these line emissions more sensitive to the precise
physical details (e.g. inhomogeneities like clumpiness) in the
immediate vicinity of the star.
Since the emission of forbidden lines depends on square of the local density,
denser clumps in an inhomogeneous medium will show enhanced emission
compared to an equivalent uniform medium.
Hence, a clumpy medium around the exciting star in \mbox{IRAS~10031-5632}, 
could help explain the 
observed higher nebular line luminosities. This scenario is also 
consistent with the higher radio continuum emission observed. 

\section{Discussion}

\subsection{IRAS~10049-5657}

Using the FIR map of \mbox{IRAS~10049-5657} (Fig.~\ref{FIRmap_a}), we 
see an extension of the dust emission towards north-east. This is also seen in 
the 60 and 100 $\micron$ emission IRAS-HIRES maps of this region, shown in 
Fig.~\ref{HIRES_a}. The IRAS-HIRES mid infrared emission from warm dust, 
however, shows an extension towards the west. This western extension is 
also clearly seen in the MSX maps. The flux densities from the TIFR maps 
at 150 and 210 $\micron$ have been used to compute the mass of dust using the 
formulation of Hildebrand (1983) and Sandell (2000). For a temperature of 30 K,
 the dust mass obtained is  $\sim21$ $M_{\odot}$. As can be seen from 
Fig.~\ref{SUMSS_a}, a north-east extension seen also in the cold dust 
emission is 
observed for the ionised gas also. The total flux of 42.4 Jy at 843 MHz can be 
compared with the flux density of 40 Jy (beam $\sim14'$) at 1410 MHz 
(Manchester, 1969). 

From the CM (J-H vs J) diagram of the 2MASS sources (Fig.~\ref{cmcc_a} left) 
within the 
cluster radius ($\sim65''$), we see that there are 11 sources lying above the 
ZAMS curve of O9. These 11 sources are designated as IRA-1, ... , IRA-11 and a 
list of their positions and magnitudes is given in Table~\ref{2MASS_parm_a}. 
However, it is important to note that the spectral types inferred from CM 
diagram are the earliest possible spectral type (upper limits) when the stars 
have infrared excess. Sources with infrared excess can be 
found from the CC (J-H vs H-K) diagram. In the CC diagram (see 
Fig.~\ref{cmcc_a}), 8 `asterisks' lie in the band occupied by reddened ZAMS 
stars. One source, IRA-7 (shown as a triangle lying in the `T' region), shows an 
infrared excess while IRA-10 lies to 
the left of the reddening band of the ZAMS objects (drawn from the top
 of the main sequence branch) in the CC diagram. The upper left part of 
the CC diagram is not an allowed region for young stellar objects (Lada 
\& Adams, 1992). IRA-10 is faint in J band 
($16.9\pm0.2$ mag) with larger errors though it is relatively 
brighter in the H ($14.2\pm0.02$ mag) and K$_s$ ($13.2\pm0.04$) bands 
as compared to the J band. One possibility to explain the near infrared 
colours of IRA-10 as well as its position in the CC diagram is that IRA-10 
could comprise two or more unresolved sources. The sources 
lying near the unreddened main sequence in the CC diagram are 
possibly foreground sources not associated with the cluster. A small but 
significant fraction of 
cluster stars lie outside and to the right of the reddening band of the ZAMS 
stars. These are mostly young stellar objects (YSOs) with intrinsic colour 
excess. Fourteen objects lie in this infrared excess zone, i.e. in the 
`T' and `P') regions. By dereddening the 
stars (on the CC diagram) that fall within the reddening vectors encompassing 
the main sequence stars, we find the visual extinction (A$_V$) towards each 
star. The individual extinction values range from $A_V \sim 0 - 17$ magnitudes.
From the CM diagram, we find that the extinction values of most 
`asterisks' lie at about $A_V\sim 15 - 18$ magnitudes. 

\begin {figure}
\epsscale{1}
\vspace* {-4cm}
\plotone{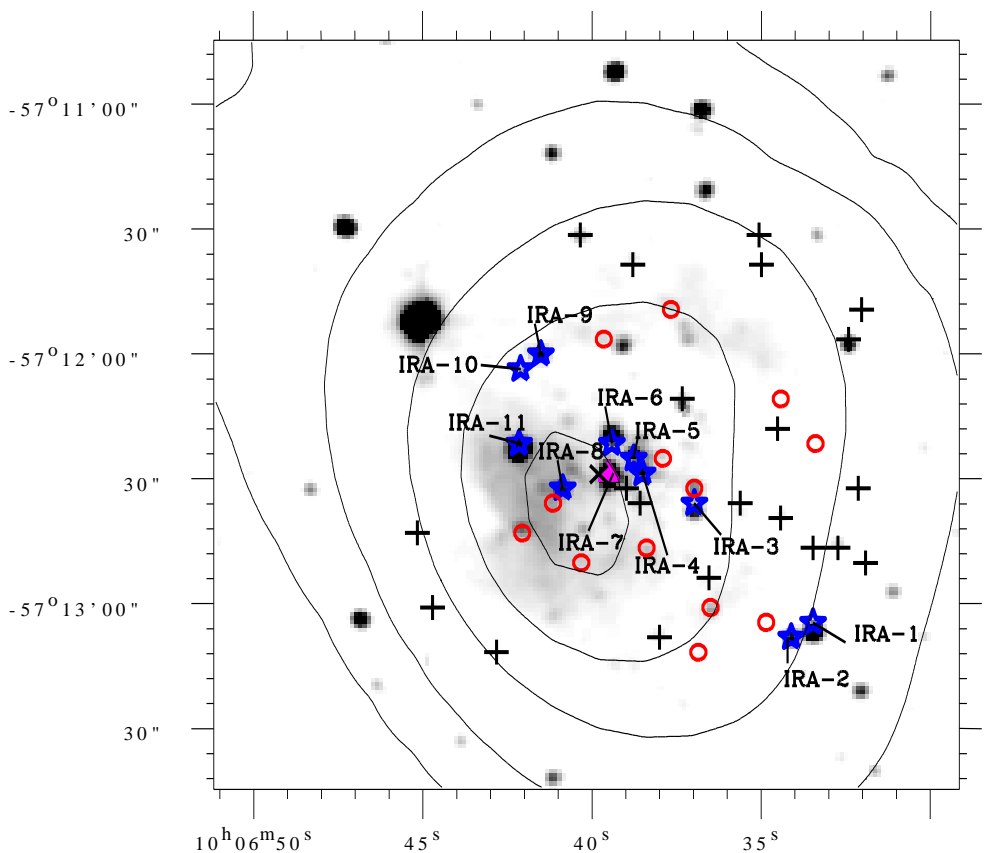}
\vspace* {-5cm}
\caption{The grayscale 2MASS K$_s$-band image for the region
around \mbox{IRAS~10049-5657}. The labelled
axes are in J2000 coordinates. Overlaid on top are the SUMSS 843 MHz radio 
contours. The location of the IRAS peak is represented by a cross. Various 
2MASS PSC sources are marked. The 
asterisk symbols represent sources lying above the ZAMS curve of spectral type 
O9. The open circles represent sources depicting an infrared excess in
the colour-colour diagram. The filled triangles represent sources lying 
above the ZAMS curve of spectral type O9 as well as having an infrared excess. 
The other sources are marked by plus symbols.}
\label{2mass_a}
\end {figure}

Figure~\ref{2mass_a} shows the 2MASS K$_s$-band image of the region around 
\mbox{IRAS~10049-5647} in grayscale. The 2MASS image of this region shows diffuse 
emission apart from the sources (earlier than spectral type O9) clustered 
together. 
It is interesting to note the clustering of these sources close to the IRAS 
peak and the distribution of the other such stars in the NE-SW direction
which is the direction of extension of the ionised gas as well as the cold 
dust. The morphology of emission in the UIB 
as well as the emission from warm dust in MSX bands also shows extension along 
the south-west direction. The brightest infrared 
source among the selected 2MASS sources is 
IRA-11. IRA-7 is closest ($\sim 2.7''$) to the IRAS peak, is of 
spectral type earlier than O5 and has an infrared excess. This is consistent 
with the spectral type determined by Bik et al. (2005) for IRA-7 (referred to 
as 10049nr411 in their paper) using high resolution K-band spectra. They find 
that it is of spectral type O3-O4. 
Recent multi-epoch radial velocity measurements by Apai et al. (2007) show 
large amplitude variations in radial velocities pointing towards IRA-7 being a 
massive binary system (50 and 20 M$_{\odot}$).
Bik et al. (2005) have also carried out spectroscopy of 
IRA-3 (10049nr324) and find it to be of spectral type 
O3-O4/O5-O6. This is consistent with its ZAMS spectral type, O5-O6, obtained 
from the CM diagram. It is interesting to note that the `asterisk' closest to 
the radio peak is IRA-8. The CM diagram indicates IRA-8 to be of ZAMS spectral 
type O9-O6. 

We have also cross-correlated the MSX PSC sources with those from
2MASS PSC which lie within the cluster radius. There are 2 such MSX PSC
sources namely G282.0341-01.1810 and G282.0176-01.1793 (hereafter M1 
and M2, respectively; listed in Table~\ref{msx}). M1 coincides ($1.3''$) with 
an infrared excess source, 2MASS-J10064115-5712377 (J$\sim16.0\pm0.2$, H$\sim14.9\pm0.2$, K$_s\sim13.5\pm0.3$). M2 coincides ($2''$) 
with a source detected only in K$_s$ band, 2MASS-J10063592-5711563 
(K$_s\sim14.1\pm0.2$ mag). From the 2MASS-MSX 
colour-colour diagram ($F_{21}/F_8$ vs $F_8/F_K$; Lumsden et al. 2002, their 
Fig.~9), we find that these sources lie in the region generally covered by 
compact H II regions and massive young stellar objects. We can, therefore, 
conclude that these are young stars associated with the cluster.

\subsection{IRAS~10031-5632}

The dust emission around \mbox{IRAS~10031-5632} is compact (unresolved) at 150 $\mu$m,
as well as at the IRAS-HIRES wavebands. \mbox{IRAS~10031-5632} is not
resolved in the SUMSS radio map also. However, \mbox{IRAS~10031-5632} is barely 
resolved in the 210 $\mu$m map. Deconvolving the beam from 210 $\mu$m image 
gives us $\sim0.9'$ as an estimate of its size. The dust mass obtained 
using the flux density at 210 $\micron$ from the TIFR map is  
$\sim7$ $M_{\odot}$ for a temperature of 30 K. However, from the radiative 
transfer modelling, we obtain a larger dust mass of 24 $M_{\odot}$. 
The total radio flux density is 1.1 Jy. Considering that the radio flux is from 
a single ionising ZAMS star, we use the formulation of Schraml \& Mezger (1969)
and Panagia (1973) to estimate its spectral type. We find it to be of ZAMS 
spectral type O9-O8.5. This can be compared with the ZAMS spectral type of O9 
obtained from the FIR luminosity. 

\begin {figure}
\epsscale{1}
\vspace* {-4cm}
\plotone{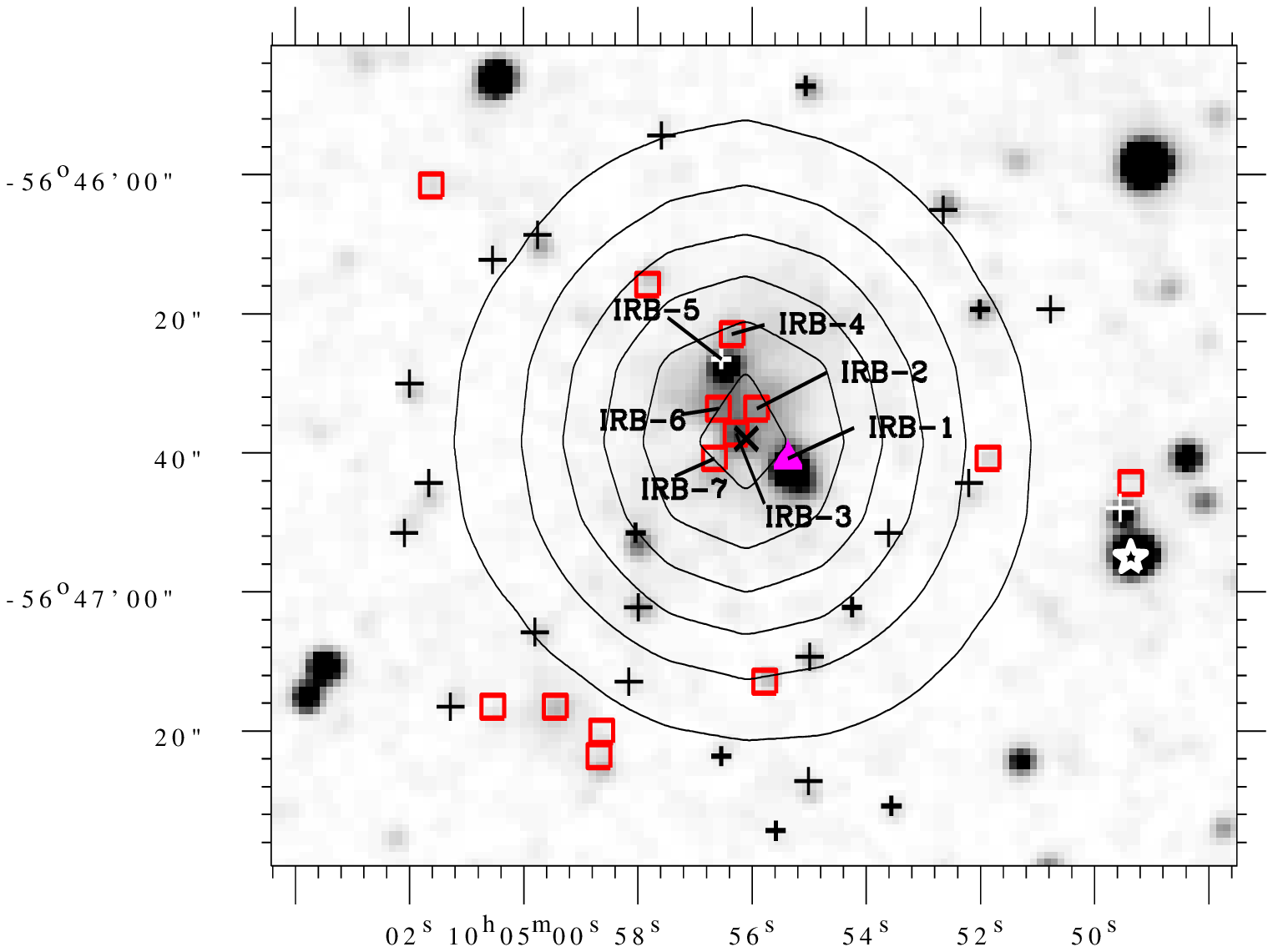}
\vspace* {-5cm}
\caption{The grayscale 2MASS K$_s$-band image for the region
around \mbox{IRAS~10031-5632}. The labelled
axes are in J2000 coordinates. Overlaid on top are the SUMSS 843 MHz radio 
contours. The location of the IRAS peak is represented by a cross. 
Various 2MASS PSC sources are marked.
The asterisk symbol represents the sources lying above the
ZAMS curve of spectral type O9. The solid triangle represents the 
 source lying above the ZAMS curve of spectral type O9 and having an infrared
excess. The open squares represent sources detected in 
either H and K$_s$ bands or only K$_s$ band. The plus symbols represent the 
other sources.} 
\label{2mass_b}
\end {figure}

Unlike the case of \mbox{IRAS~10049-5657}, we do not see a rich cluster with a large 
number of stars around \mbox{IRAS~10031-5632}. Rather a small group 
comprising of a few stars is seen close to this region. There are seven 
sources (labelled as IRB-1 to IRB-7) clustered close to the 
FIR (IRAS) peak; of which two, IRB-1 and IRB-5, are detected in all the 
three (J, H and K$_s$) bands. The 2MASS designation and flux details of these 
sources are given Table~\ref{2MASS_parm_b}. From the CM diagram (H-K vs K; 
Fig.~\ref{cmcc_b} left), it is observed that IRB-1 is of ZAMS spectral type 
earlier than O5. It is to be noted that this spectral type of IRB-1 obtained 
from NIR study is the earlier possible spectral type 
(upper limit) as it shows an infrared excess. IRB-5 is however of 
later spectral type, B3-B2. It lies among a well-defined group of stellar 
sources which have 
lower extinction values (A$_V\sim5$ mag). From the CC diagram (J-H vs H-K), 
it is found to lie near the main-sequence curve. It is therefore possible that 
IRB-5 is not associated with the cluster but is a foreground source. 
The other five sources are detected in H and K$_s$ or only in K$_s$ 
bands. The 
source closest ($\sim 1.5''$) to the IRAS peak, IRB-3, is detected in H 
($15.1\pm0.2$ mag) and K$_s$ ($12.8\pm0.1$ mag) 
bands but not detected in the J band. From the CM diagram, it is found that 
IRB-3 is a heavily extincted early spectral type (O9-O6) ZAMS star. It 
is located at the peak of 843 MHz radio emission. The other 
four sources IRB-2, IRB-4, IRB-6 and IRB-7 are detected only in K$_s$ band. 
This implies that these are deeply embedded objects. Thus, among the 
small number of cluster members, the majority of them are deeply 
embedded indicating that the star formation here is in an early stage. 
The 2MASS image of the region around \mbox{IRAS~10031-5632} is shown in 
Fig.~\ref{2mass_b}. The sources lying above the ZAMS curve of spectral type O9 are 
shown as asterisk symbols while the `asterisk' having an infrared excess
is shown as a solid triangle. The open squares represent sources detected  
either in H and K$_s$ bands or only in K$_s$ band. The other 
sources are represented with plus-symbols. The seven sources discussed 
above are labelled as ``IRB" in the 
figure. It is interesting to note the presence of young objects (detected 
in H and K$_s$ or only K$_s$ bands) clustered near the IRAS peak. 

In the \mbox{IRAS~10031-5632} region, we find that there is 1 MSX PSC source within 
$1'$ radius centered on the IRAS peak, G281.5857-00.9706 (designated as M3 and  
listed in Table~\ref{msx}). M3 could possibly be associated with either IRB-3 
($\sim2.8''$) or IRB-2 ($\sim2.8''$). In both the cases, we obtain the
2MASS-MSX colours ($F_{21}/F_8$ and $F_8/F_K$, Lumsden et al. 2002) and
observe that M3 lies in the general region covered by compact H II regions. 

\section{Summary}

The massive star forming regions associated with \mbox{IRAS~10049-5657} and 
\mbox{IRAS~10031-5632} have been studied using the infrared (near, mid and far) wavebands.
The dust and gas environments as well as the stellar sources have been probed 
using data from TIFR balloon-borne telescope, MSX, SUMSS and 2MASS. The 
spatial distributions of far infrared emission from cold dust at 150 and 210 
$\mu$m have been obtained alongwith maps of optical depth ($\tau_{200}$) and 
colour temperature, T(150/210), Using MSX data, the emission from warm dust 
and UIBs in the \mbox{IRAS~10049-5657} region has been studied. 
\mbox{IRAS~10049-5657} region shows the presence of a rich cluster of OB stars 
which gives rise to strong radio continuum ($\sim42$ 
Jy at 843 MHz). The cluster radius is estimated to be $\sim65''$ and within the 
cluster radius, there are 11 2MASS sources lying above the ZAMS curve of 
spectral type O9, designated IRA-1 to IRA-11. One source, IRA-7 also has an 
infrared excess and is located closest to the IRAS peak. It is a massive star 
of early spectral type, earlier than O5 consistent with earlier studies. 
Unlike \mbox{IRAS~10049-5657}, the star forming 
region associated with \mbox{IRAS~10031-5632}, comprises of a very 
compact group of stars. The SUMSS radio flux at 843 MHz predicts a radio 
spectral type of O9-O8.5. The sources clustered near the IRAS peak comprise 
of young sources detected in H and K$_s$ or only K$_s$ bands. The 2MASS source
 detected in all the three bands, designated IRB-1, has an infrared excess and 
is found to be of ZAMS spectral type earlier than O5 from the 2MASS 
colour-magnitude diagram. Self-consistent radiative transfer modelling 
constrained by
 observations have been carried out for both these sources. The geometric 
details of the clouds, the dust-composition and optical depths, etc have 
been obtained from the best fit models. We have also carried out modelling of 
line emission from \mbox{IRAS~10049-5657} and \mbox{IRAS~10031-5632} using a 
scheme based on 
CLOUDY. The predictions of the model for emission in
 the ionic lines are closer to LRS detections for \mbox{IRAS~10049-5657}
 than for \mbox{IRAS~10031-5632}. We speculate that for \mbox{IRAS~10031-5632},
 some of the 
basic assumptions of the modelling scheme are not valid.

\textit{Acknowledgments}

We thank the anonymous referee for useful suggestions that improved
the paper.
It is a pleasure to thank the members of the Infrared Astronomy Group at 
TIFR for their support during laboratory tests and balloon flight campaigns. 
All members of the Balloon Group and Control Instrumentation Group of the TIFR 
Balloon Facility, Hyderabad, are thanked for their technical support 
during the flight. We thank IPAC, Caltech, for providing us the 
HIRES-processed IRAS products.




\clearpage

\begin{table}
\caption{Flux density details of \mbox{IRAS~10049-5657} and 
\mbox{IRAS~10031-5632}}
\vspace{1cm}
\label{fluxes}
\begin{tabular}{|c |c c| c c c c |c c c c|} \hline \hline
 IRAS PSC & \multicolumn{10}{c|}{Flux Density (Jy) for $\lambda$ ($\micron$)} \\ \hline
 Source & \multicolumn{2}{c|}{TIFR images$^a$} & \multicolumn{4}{c|}{IRAS-HIRES
images$^a$} & \multicolumn{4}{c|}{MSX images$^a$} \\
 & \multicolumn{2}{c|}{} & \multicolumn{4}{c|}{IRAS PSC} & \multicolumn{4}{c|}{ } \\ \hline
 & 210 & 150 & 100  & 60 & 25 & 12 & 21.3 & 14.7 & 12.1 & 8.3 \\ \hline
 10049-5657 & 1147 & 2534 & 6224 & 8901 & 2166 &
368 & 1486 & 642 & 373 & 120 \\
 &     &      & 7023 & 6146 & 780 & 148 &  &  &  &  \\
 &     &      &      &      &     &     &  &  &  &   \\
 10031-5632 & 679 & 1405 & 1574 & 1381 & 215 & 25 & 126 & 23 & 22 & 15 \\
 &     &      & 1512 & 1166 & 209 & 168 &  &  &  &  \\ \hline
\end{tabular}

$^a$ Fluxes obtained by integrating over a circular region of diameter $3\arcmin$ centred on the peak. \\
\end{table}


\begin{table*}
\begin{center}
\caption{Best-fit parameters of the radiative transfer models for 
\mbox{IRAS~10049-5657} and \mbox{IRAS~10031-5632}.}
\vspace*{1cm}
\hspace*{-1.5cm}
\label{radtran_parm}
\begin{tabular}{|c c c c c c c c c c|} \hline \hline
Source & $\alpha$ & $R_{max}$ & $R_{min}$ & $r_{H II}$ & $\tau_{100}$ &
L$^a$ & Dust & Composition & M$_{dust}$ \\
IRAS   &          &   (pc)    &   (pc)    &   (pc)    &              &
(10$^5$ $L_\odot$)  &  Type & Silicate:Graphite   &  ($M_\odot$) \\ \hline
10049-5657 & 0.0  & 4.2 & 0.009 & 2.8 & 0.002 &  10 & DL & 11:89 & 12 \\
10031-5632 & 0.0 & 1.0 & 0.007 & 0.03 & 0.07 & 0.42 & DL & 62:38 & 24 \\ \hline
\end{tabular}

$^a$ Luminosity of single ZAMS star used in the model
\end{center}
\end{table*}



\begin{table}
\begin{center}
\caption
{Emergent line luminosities predicted by the model for \mbox{IRAS~10049-5657}. 
The line-to-continuum ratios for the model (convolved with the IRAS-LRS 
instrument profile) are also listed. Also 
included are the line luminosities and the ratio of line-to-continuum for the 
lines detected in the IRAS-LRS spectrum of \mbox{IRAS~10049-5657}.}
\label{model_linea}
\vspace*{0.5cm}
\begin{tabular}{l r r r r r} \hline \hline
 & & \multicolumn{2}{c}{Model} & \multicolumn{2}{c}{Observed (LRS)} \\ \hline
Element & Wavelength & Luminosity & $\frac{L(line)}{L(cont)}$ & Luminosity & $\frac{L(line)}{L(cont)}$ \\
Ionisation Stage & ($\mu$m) & ($L_{\odot}$) &  & ($L_{\odot}$) & \\ \hline
$[$C II$]$ & 157.78 & 71.54 & & - & - \\
$[$O I$]$  & 145.63 & 1.21 & & - & - \\
$[$N II$]$ & 121.84 & 19.53 & & - & -  \\
$[$O III$]$ & 88.42 & 2577.02 & & - & - \\
$[$O I$]$ & 63.23 & 13.50 & & - & - \\
$[$N III$]$ & 57.26 & 772.85 &  & - & - \\
$[$O III$]$ & 51.85 & 2924.28 & & - & - \\
$[$Ne III$]$ & 36.04 & 89.82 &  & - & - \\
$[$Si II$]$ & 34.84 & 87.99 & & - & - \\
$[$S III$]$ & 33.50 & 1052.20 &  & - & - \\
$[$O IV$]$ & 25.91 & 114.36 & & -  & - \\
$[$Ar III$]$ & 21.84 & 20.60 & 0.002 & 263 & 0.04 \\
$[$S III$]$ & 18.69 & 765.01 & 0.08 & 300 & 0.03 \\
$[$Ne III$]$ & 15.57 & 994.78 & 0.15 & 386 & 0.04 \\
$[$Ar V$]$  & 13.11 & 1.02 & 0.001 & - & - \\
$[$Ne II$]$ & 12.82 & 62.14 & 0.03 & 250 & 0.06 \\
$[$S IV$]$ & 10.52 & 1101.83 & 0.35 & 35 & 0.01 \\
$[$Ar III$]$ & 8.99 & 289.82 & 0.10 & - & - \\
$[$Ar V$]$  & 7.91 & 0.63 & & - & - \\
$[$Ar II$]$ & 7.00 & 9.82 & & - & - \\ 
$[$Ar VI$]$ & 4.53 & 0.21 & & - & - \\ \hline \hline
\end{tabular}
\end{center}
\end{table}



\begin{table}
\begin{center}
\caption
{Emergent line luminosities predicted by the model for \mbox{IRAS~10031-5632}. 
The line-to-continuum ratios for the model (convolved with the IRAS-LRS 
instrument profile) are also listed. Also
included are the line luminosities and the ratio of line-to-continuum for the
lines detected in the IRAS-LRS spectrum of \mbox{IRAS~10031-5632}.}
\label{model_lineb}
\vspace*{0.5cm}
\begin{tabular}{l r r r r r} \hline \hline
 & & \multicolumn{2}{c}{Model} & \multicolumn{2}{c}{Observed (LRS)} \\ \hline
Element & Wavelength & Luminosity & $\frac{L(line)}{L([cont])}$ & Luminosity & $\frac{L(line)}{L([cont])}$ \\
Ionisation Stage & ($\mu$m) & ($L_{\odot}$) &  & ($L_{\odot}$) & \\ \hline
$[$C II$]$ & 157.78 & 0.31 & & - & - \\
$[$O I$]$  & 145.63 & 0.09 & & - & - \\
$[$N II$]$ & 121.84 & 0.01 & & - & - \\
$[$O III$]$ & 88.42 & 0.40 & & - & - \\
$[$O I$]$ & 63.23 &  0.56 & & - & - \\
$[$N III$]$ & 57.26 & 0.36 & & - & - \\
$[$O III$]$ & 51.85 & 2.90 & & - & - \\
$[$Ne III$]$ & 36.04 & 0.33 & & - & - \\
$[$Si II$]$ & 34.84 & 0.08 & & - & - \\
$[$S III$]$ & 33.50 & 0.53 & & - & - \\
$[$Ar III$]$ & 21.84 & 0.10 & 0.0002 & - & -\\
$[$S III$]$ & 18.69 & 1.76 & 0.0025 & 16.6 & 0.07 \\
$[$Ne III$]$ & 15.57 & 3.13 & 0.004 & - & - \\
$[$Ne II$]$ & 12.82 & 1.61 & 0.005 & 30 & 0.5 \\
$[$S IV$]$ & 10.52 & 0.27 & 0.0006 & - & - \\
$[$Ar III$]$ & 8.99 & 0.40 & 0.0009 & 9.3 & 0.12 \\
$[$Ar II$]$ & 7.00 & 0.17 & & - & - \\ \hline \hline 
\end{tabular}
\end{center}
\end{table}



\begin{table*}
\begin{center}
\caption{Details of 2MASS PSC sources earlier than O9 around 
\mbox{IRAS~10049-5657}.}
\vspace*{1cm}
\hspace*{-1.5cm}
\label{2MASS_parm_a}
\begin{tabular}{|c c c c c c c |} \hline \hline
2MASS PSC & Name$^a$ & $\alpha_{2000}$ & $\delta_{2000}$ & J & H & K$_s$\\
designation & & (deg) & (deg) & (mag) & (mag) & (mag) \\ \hline
J10063343-5713064 & IRA-1 & 151.639294 & -57.218468 & $13.18\pm0.05$ & $11.41\pm0.05$ & $10.66\pm0.04$ \\
J10063408-5713079 & IRA-2 & 151.642008 & -57.218887 & $13.57\pm0.02$ & $12.28\pm0.02$ & $11.79\pm0.02$ \\
J10063696-5712372 & IRA-3 & 151.654013 & -57.210342 & $14.46\pm0.03$ & $12.72\pm0.03$ & $11.84\pm0.02$ \\
J10063849-5712287 & IRA-4 & 151.660405 & -57.207985 & $14.57\pm0.08$ & $12.90\pm0.12$ & $11.92\pm0.06$ \\
J10063875-5712243 & IRA-5 & 151.661484 & -57.206776 & $14.56\pm0.09$ & $12.75\pm0.12$ & $11.80\pm0.10$ \\ 
J10063940-5712198 & IRA-6 & 151.664190 & -57.205513 & $13.79\pm0.03$ & $11.97\pm0.05$ & $10.95\pm0.04$ \\     
J10063949-5712299 & IRA-7 & 151.664564 & -57.208324 & $12.89\pm0.04$ & $11.44\pm0.06$ & $10.36\pm0.04$ \\
J10064088-5712327 & IRA-8 & 151.670346 & -57.209087 & $14.97\pm0.09$ & $13.20\pm0.08$ & $12.13\pm0.10$ \\
J10064153-5712003 & IRA-9 & 151.673054 & -57.200096 & $14.77\pm0.03$ & $12.90\pm0.03$ & $11.89\pm0.04$ \\
J10064211-5712041 & IRA-10 & 151.675482 & -57.201160 & $16.88\pm0.18$ & $14.21\pm0.02$ & $13.17\pm0.04$ \\
J10064217-5712231 & IRA-11 & 151.675725 & -57.206444 & $11.33\pm0.02$ & $9.96\pm0.03$  & $9.38\pm0.03$ \\
\hline
\end{tabular}
\end{center}

$^a$ Short name used in the present work \\
\end{table*}


\begin{landscape}
\begin{table*}
\begin{center}
\caption{Details of the MSX PSC sources close to \mbox{IRAS~10049-5657} and 
\mbox{IRAS~10031-5632} (see text for details).}
\vspace*{1cm}
\hspace*{-1.5cm}
\label{msx}
\begin{tabular}{|c c c c c c c c c|} \hline \hline
IRAS & MSX PSC & Name$^a$ & $\alpha_{2000}$ & $\delta_{2000}$ & $F_{8}$ & $F_{12}$ &
$F_{14}$ & $F_{21}$\\
Association & designation & & (deg) & (deg) & (Jy) & (Jy) & (Jy) & (Jy) \\ \hline
IRAS~10049-5657 &  G282.0341-01.1810 & M1 & 151.6721 & -57.2105 & $16.5\pm0.7$ & $56.9\pm2.8$ & $97.0\pm5.9$ & $195.1\pm11.7$ \\
IRAS~10049-5657 &  G282.0176-01.1793 & M2 & 151.6493 & -57.1995 & $3.2\pm0.1$ & $13.1\pm0.7$ & $23.5\pm1.4$ & $39.2\pm2.3$ \\
 &  &  &  &  &  &  &  &  \\
 IRAS~10031-5632 & G281.5857-00.9706 & M3 & 151.2342 & -56.7765 & $8.0\pm0.3$ & $14.7\pm0.7$ & $17.1\pm1.0$ & $93.2\pm5.6$ \\ \hline
\end{tabular}
\end{center}

$^a$ Short name used in the present work \\
\end{table*}
\end{landscape}



\begin{table*}
\begin{center}
\caption{Details of 2MASS PSC sources close to IRAS~10031-5632.}
\vspace*{1cm}
\hspace*{-1.5cm}
\label{2MASS_parm_b}
\begin{tabular}{|c c c c c c c |} \hline \hline
2MASS PSC & Name$^a$ & $\alpha_{2000}$ & $\delta_{2000}$ & J & H & K$_s$\\
designation & & (deg) & (deg) & (mag) & (mag) & (mag) \\ \hline
J10045536-5646422 & IRB-1 & 151.230694 & -56.778397 & $15.25\pm0.09$ & $12.56\pm0.07$ & $10.70\pm0.04$ \\
J10045591-5646340 & IRB-2 & 151.232992 & -56.776119 & $-$ & $-$ & $13.24\pm0.20$ \\
J10045627-5646381 & IRB-3 & 151.234494 & -56.777267 & $-$ & $15.14\pm0.24$ & $12.83\pm0.13$ \\
J10045634-5646226 & IRB-4 & 151.234762 & -56.772972 & $-$ & $-$ & $13.80\pm0.18$ \\
J10045653-5646275 & IRB-5 & 151.235575 & -56.774307 & $12.14\pm0.03$ & $11.85\pm0.04$ & $11.69\pm0.06$ \\
J10045658-5646324 & IRB-6 & 151.235780 & -56.775673 & $-$ & $-$ & $13.11\pm0.13$ \\
J10045666-5646424 & IRB-7 & 151.236113 & -56.778446 & $-$ & $-$ & $14.02\pm0.11$ \\
\hline
\end{tabular}
\end{center}

$^a$ Short name used in the present work \\
\end{table*}


\clearpage

\newpage

\newpage


\begin{thebibliography}{}

\bibitem[2007]{Ap07}Apai, D., Bik, A., Kaper, L., Henning, T. \& Zinnecker, H. 
2007, ApJ, 655, 484 
\bibitem[1990]{Au90} Aumann, H. H., Fowler, J. W. \& Melnyk, M. 1990, AJ, 99,
1674
\bibitem[1988]{Be88} Bessel, M. S. \& Brett, J. M. 1988, PASP, 100, 1134
\bibitem[2005]{Bi05} Bik, A., Kaper, L., Hanson, M. M. \& Smits, M. 2005, A\&A, 440, 121
\bibitem[2001]{Bl01} Blum, R. D., Damineli, A. \& Conti, P. S. 2001, AJ, 121, 3149
\bibitem[1999]{Bo99} Bock, D. C., Large, M. I. \& Sadler, E. M. 1999, AJ, 117, 1578
\bibitem[1989]{Br89} Braz, M. A. Gregorio Hetem, J. C., Scalise, E., Jr., Monteiro Do Vale, J. L. \& Gaylard, M. 1989, A\&AS, 77, 465
\bibitem[1996]{Br96} Bronfman, L., Nyman, L.-A. \& May, J. 1996, A\&AS, 115, 81
\bibitem[1987]{Ca87} Caswell, J. L. \& Haynes, R. F. 1987, A\&A, 171, 261
\bibitem[1995]{Co95} Cohen, R. J., Masheder, M. R. W. \& Caswell, J. L. 1995, MNRAS, 274, 808
\bibitem[2004]{Co04} Conti, P. S. \& Crowther, P. A. 2004, MNRAS, 355, 899
\bibitem[1990]{de90} de Muizon, J. M., D'Hendecourt, L. B. \& Geballe, T.
R. 1990, A\&A, 235, 367
\bibitem[1984]{Dr84} Draine, B. T. \& Lee, H. M. 1984, ApJ, 285, 89
\bibitem[2003]{Eg03} Egan, M. P., Price, S. D., Kraemer, et al.
2003, Air Force Research Laboratory Technical Report, AFRL-VS-TR-2003-1589
\bibitem[1984]{Ga84} Gardner, F. F. \& Whiteoak, J. B. 1984, MNRAS, 210, 23
\bibitem[1988]{Gh88} Ghosh, S. K., Iyengar, K. V. K., Rengarajan, T. N., Tandon, S. N., Verma, R. P. \& Daniel, R. R. 1988, ApJ, 330, 928
\bibitem[2000]{Gh00} Ghosh, S. K., Mookerjea, B., Rengarajan, T. N., Tandon, 
S. N. \& Verma, R. P. 2000, A\&A, 363, 744
\bibitem[2002]{Gh02} Ghosh, S. K. \& Ojha, D. K. 2002, A\&A, 288, 326
\bibitem[1978]{Gu78} Gull, S. F. \& Daniell, G. J. 1978, Nature, 272, 686
\bibitem[2003]{Ha03} Hanson, M. M., Kaper, L., Bik, A. et al. 2003, IAUS, 212, 
467 
\bibitem[1983]{Hi83} Hildebrand, R. H. 1983, QJRAS, 24, 267
\bibitem[1968]{Hi68} Hill, E. R. 1968, AuJP, 21, 746
\bibitem[1986]{IR86} IRAS Science Team, 1986, A\&AS, 65, 607
\bibitem[2001]{Ka01} Karnik, A. D., Ghosh, S. K., Rengarajan, T. N. \& Verma, 
R. P. 2001, MNRAS, 326, 293
\bibitem[1997]{Ku97} Kuchar, T. A. \& Clark, F. O. 1997, ApJ, 488, 224
\bibitem[1992]{La92} Lada, C. J. \& Adams, F. C. 1992, ApJ, 393, 278
\bibitem[2002]{Lu02} Lumsden, S. L., Hoare, M. G., Oudmaijer, R. D. \&
Richards, D. 2002, MNRAS, 336, 621
\bibitem[1969]{Ma69} Manchester, B. A. 1969, AuJP, 12, 13
\bibitem[1997]{Me97} Meyer, M. R., Calvet, N. \& Hillenbrand, L. A. 1997, AJ,
114, 288
\bibitem[1982]{Ma82} Mezger, P. G., Mathis, J. S. \& Panagia, N. 1982, A\&A,
105, 372
\bibitem[1999]{Mo99} Mookerjea, B. \& Ghosh, S. K. 1999, J. Astrophys. Astr.
20, 1
\bibitem[2002]{Oj02} Ojha, D. K., Ghosh, S. K. \& Verma R. P. 2002, Bulletin 
of Astronomical Society of India, 30, 647
\bibitem[2004]{Oj04} Ojha, D.K., Tamura, M., Nakajima, Y. et al. 2004, 
ApJ, 608, 797
\bibitem[1973]{Pa73} Panagia, N. 1973, AJ, 78, 929
\bibitem[1992]{Pe92} Peng, R. S. \& Whiteoak, J. B. 1992, MNRAS, 254, 301
\bibitem[2001]{Pr01} Price, S. D., Egan, M. P., Carey, S. J.,
Mizuno, D. R., \& Kuchar, T. A. 2001, AJ, 121, 2819
\bibitem[2002]{Pu02} Puchalla, J. L., Caldwell, R., Cruz, K. L., Devlin, M. J., Dorwart, W. B., et al. 2002, AJ, 123, 1978
\bibitem[1985]{Ri85} Rieke, G. H. \& Lebofsky, M. J. 1985, ApJ, 288, 618
\bibitem[2000]{Sa00} Sandell, G. 2000, A\&A, 358, 242
\bibitem[1993]{Sc93} Schutte, A. J., van der Walt, D. J., Gaylard, M. J. \& 
MacLeod, G. C. 1993, MNRAS, 261, 783
\bibitem[1969]{Sc69} Schraml, J. \& Mezger, P. G. 1969, ApJ, 156, 269 
\bibitem[1990]{Si90} Simpson, J. P. \& Rubin, R. H. 1990, ApJ, 354, 165
\bibitem[1994]{Ve94} Verma, R. P., Bisht, R. S., Ghosh, S. K. et al. 1994, A\&A, 284, 936
\bibitem[2007]{Vi07} Vig, S., Ghosh, S. K., Ojha, D. K. \& Verma, R. P. 2007, 
A\&A, 463, 175
\bibitem[2000]{Vi00} Vilas-Boas, J. \& Abraham, Z. 2000, A\&A, 355, 1115
\bibitem[1989]{Vo89} Volk, K. \& Cohen, M. 1989, AJ, 98, 931
\bibitem[1974]{Wh74} Whiteoak, J. B. \& Gardner, F. F. 1974, A\&A, 37, 389
\bibitem[1982]{Wh82} Whiteoak, J. B., Otrupcek, R. E., Rennie, C. J. 1982, PASAu, 4, 434
\bibitem[1992]{Za92} Zavagno, A., Cox, P. \& Baluteau, J. -P. 1992, A\&A, 259, 241

\end{thebibliography}
\end{document}